\documentclass[12pt]{article}
\usepackage{amssymb,graphicx}
\usepackage{epstopdf}
\usepackage{amsmath,amsfonts}
\usepackage{epsfig}
\begin{document}

\sloppy

\title{\bf Constraining\\ anisotropic models of the early universe\\ with WMAP9 data}
\author{S.~R.~Ramazanov$^{a}$\footnote{{\bf e-mail}: Sabir.Ramazanov@ulb.ac.be} 
G.~I.~Rubtsov$^{b,c,d}$\footnote{{\bf e-mail}: grisha@ms2.inr.ac.ru}
\\
$^a$ \small{\em Universit\'e Libre de Bruxelles,}\\
\small{\em Service de Physique Th\'eorique, CP225, Boulevard du Triomphe, B-1050 Brussels, Belgium}\\ 
$^b$ \small{\em Institute for Nuclear Research of the Russian Academy of Sciences,}\\
\small{\em Prospect of the 60th Anniversary of October 7a, 117312 Moscow, Russia}\\
$^c$ \small{\em Physics Department, Moscow State University,} \\
\small{\em Vorobjevy Gori 1, 119991 Moscow, Russia}\\
$^d$ \small{\em Novosibirsk State University,}\\
\small{\em Pirogov street 2, 630090 Novosibirsk, Russia}
} 
\maketitle

\begin{abstract}
We constrain several models of the early Universe that predict a 
statistical anisotropy of the cosmic microwave background (CMB) sky. We make use 
of WMAP9 maps deconvolved with beam asymmetries. As compared to previous releases of WMAP data, they 
do not exhibit the anomalously large quadrupole of statistical anisotropy. 
This allows to strengthen the limits on the parameters 
of models established earlier in the literature. In particular, the amplitude of the 
special quadrupole 
is constrained as $|g_*|<0.072$ at $95\%$ C.L. ($-0.046<g_*<0.048$ at $68\%$ C.L.) independently of the preferred direction 
in the sky. The upper limit is obtained 
on the total number of $e$-folds in anisotropic inflation with the Maxwellian term nonminimally coupled to the inflaton, namely 
$N_{\text{tot}} <N_{\text{CMB}} +82$ at $95\%$ C.L. ($+14$ at $68\%$ C.L.) for $N_{\text{CMB}}=60$. We also constrain models of the 
(pseudo)conformal universe. The strongest constraint is obtained 
for spectator scenarios involving a long stage of subhorizon evolution after conformal rolling, 
which reads $h^2 < 0.006$ at $95\%$ C.L., in terms of the relevant parameter. The analogous constraint is much 
weaker in dynamical models, e.g., Galilean genesis.
\end{abstract}

\section{Introduction}

The statistical isotropy of the cosmic microwave background (CMB) is one of the central pillars 
in modern cosmology. Manifesting the direction independence of the 
primordial spectrum at the onset of the hot era, it is renowned for its robustness 
against modifications in the bulk of inflationary models. Simple reasoning usually 
invokes the cosmic no-hair conjecture~\cite{Wald:1983ky} which states the rapid isotropization of the 
Universe in the presence of a positive constant energy density. 
Scalar fluctuations (e.g., carried by the inflaton) evolve in the rotationally invariant metric and 
acquire a direction-independent 
spectrum, giving rise to the statistical isotropy of CMB temperature fluctuations. 
Clearly, a violation of this property observed in the CMB sky 
would indicate a nontrivial extension of currently conventional cosmology.

One way to break statistical isotropy is by introducing vector fields with 
nonvanishing vacuum expectation values~\cite{Ackerman:2007nb, 
Golovnev:2008cf,Yokoyama:2008xw, Dimopoulos:2008yv, Abolhasani:2013bpa}. 
For example, in the Ackermann--Carroll--Wise (ACW) model~\cite{Ackerman:2007nb}, 
the idea is to add a massive vector with a fixed space-like norm. This allows 
for the anisotropic evolution of the Universe which introduces the direction dependence 
into the power spectrum of the primordial curvature perturbation $\zeta$. 
For future convenience, we write it in the generic form, 
\begin{equation}
\label{powerangen}
{\cal P}_{\zeta} ({\bf k}) ={\cal P}_{\zeta} (k) \left(1+a(k)\sum_{LM} q_{LM} Y_{LM} (\hat{{\bf k}}) \right) \; .
\end{equation}
Here $\hat{{\bf k}}$ is the direction associated with the 
cosmological wavevector ${\bf k}$; the $Y_{LM}$'s are spherical harmonics, the $q_{LM}$'s are coefficients parametrizing the 
statistical anisotropy, and $a(k)$ is the direction-independent amplitude. 
In the ACW model, the amplitude $a(k)$ is constant, i.e., it can be tuned to unity 
without loss of generality. Furthermore, the predicted statistical anisotropy is of the quadrupole type, i.e., 
only the coefficients $q_{2M}$ survive, which are not independent. In particular, 
by an appropriate choice of the reference frame, one can tune all
coefficients $q_{2M}$, except for $q_{20}$, to zero. 
The $z$ axis of this reference frame is associated with the preferred direction in the sky. 
Hereafter, we use the term {\it special quadrupole} for this type of statistical anisotropy.

Soon afterwards, it was realized that the ACW model is unstable~\cite{Himmetoglu:2008hx}. This is due to the 
longitudinal component of the massive vector that propagates as a ghost in the inflationary background. 
Similarly, all vector models of inflation violating gauge $U(1)$ symmetry share the same problem~\cite{Himmetoglu:2008hx}. 
This problem does not arise in the models of Ref.~\cite{Watanabe:2009ct} that introduce 
the Maxwellian term modified by the explicit coupling to the inflaton. 
Remarkably, for quite a wide range of coupling functions and inflaton potentials, 
one can achieve the anisotropic expansion 
of the Universe~\cite{Watanabe:2009ct} 
in a ghost free manner~\cite{Gumrukcuoglu:2010yc}. (See also Refs.~\cite{Soda:2012zm, Maleknejad:2012fw} for reviews.) 
The outcome of anisotropic inflation is still the direction dependence 
of the ACW type, while the preferred direction is associated with the electric component 
of the electromagnetic field. Interestingly, the amplitude of the special quadrupole relies on the 
overall number of $e$-folds during inflation~\cite{Bartolo:2012sd}. (See also Ref.~\cite{Lyth:2013sha}.) 
In turn, this means that the duration of inflation in those models can be strongly constrained 
with the CMB data.  

 Statistical anisotropy may follow naturally from alternative frameworks, e.g., the 
(pseudo)conformal universe~\cite{Rubakov:2009np,Creminelli:2010ba,Hinterbichler:2011qk}.  
Minimal models of this type incorporate at least two fields: 
one---which we call $\rho$---with the conformal weight $\Delta \neq 0$, and the zero-weighted conformal 
field $\sigma$. In particular, the conformal rolling scenario~\cite{Rubakov:2009np} and Galilean genesis~\cite{Creminelli:2010ba}---
two known incarnations of the pseudo(conformal) universe so far---deal with the weight $\Delta =1$ 
field $\rho$. We focus on this case in the present paper. The field $\rho$ is then assumed to have the 
time-dependent solution  
\begin{equation}
\label{backg}
\rho_0 =\frac{1}{h(t_* -t)} \; ,
\end{equation}
spontaneously breaking conformal group $SO(4,2)$ down to the de Sitter subgroup $SO(4,1)$. 
The constant $h$ is the only relevant parameter of the conformal
rolling scenario or Galilean genesis. The symmetry breaking pattern $SO(4,2) \rightarrow SO(4,1)$ 
fixes the phenomenological properties of the weight-0 
field perturbations $\delta \sigma$ 
evolving in the background~\eqref{backg} created by the field $\rho$. 
In particular---regardless of the precise details of the microscopic 
physics---one ends up with the scale-invariant power spectrum of the 
perturbations $\delta \sigma$~\cite{Hinterbichler:2011qk}. Thus, they may serve as the source of primordial fluctuations in the 
standard matter at the onset of the RD stage. 

The setup of the (pseudo)conformal universe leads unavoidably to a 
nonzero statistical anisotropy~\cite{Libanov:2010nk, Libanov:2011hh, Creminelli:2012qr}. 
This originates from the interaction between 
weight-0 and weight-1 perturbations. There are two possible types of predictions 
depending on the state of cosmological perturbations at times 
when conformal symmetry becomes irrelevant. If they are already superhorizon 
at these times, the resulting direction dependence is of the quadrupole type 
akin to anisotropic inflation~\cite{Libanov:2010nk, Creminelli:2012qr}. 
The structure of the statistical anisotropy is particularly rich in the situation 
with subhorizon cosmological perturbations~\cite{Libanov:2011hh}. 
In that case, all the coefficients $q_{LM}$ with even 
$L$ are generically nonzero in Eq.~\eqref{powerangen}. 
In both cases, the direction dependence is governed by the constant $h$. This 
gives a simple idea of how to constrain the models of interest from the 
nonobservation of statistical anisotropy in the CMB sky.

Our main goal in the present paper is to constrain models of the
early Universe that predict statistical anisotropy. We do this
in the same manner as in our previous paper~\cite{Ramazanov:2012za},
where the conformal rolling scenario was constrained by making use of
the WMAP7 data. Namely, we apply quadratic maximum likelihood 
(QML)-based estimators~\cite{Hanson:2009gu} to the CMB data and establish 
upper limits on model parameters. In the present paper,
we turn to the WMAP9 maps. With the latter, there is a strong reason
to anticipate much tighter constraints. We expect that the WMAP9 data lack
the anomalously large quadrupolar statistical anisotropy observed in
the $W$ band of the 5-year release~\cite{Groeneboom:2008fz, Groeneboom:2009cb}. The
anomaly was first interpreted as a hint towards the ACW
model. However, the preferred direction of the signal was found to be
highly aligned with poles of the ecliptic
plane~\cite{Hanson:2009gu, Groeneboom:2008fz, Groeneboom:2009cb}. Its frequency dependence 
was also suspicious: the signal so prominent in the $W$ band was
much weaker in the $V$ band~\cite{Hanson:2009gu}. Finally, results
from alternative searches, i.e., large scale structure surveys, favor
statistically isotropic primordial perturbations~\cite{Pullen:2010zy}. 
These three points strongly indicate the systematic origin of the signal
detected. In Ref.~\cite{Hanson:2009gu}, it was suggested that
beam asymmetries not accounted for in the previous analysis could
strongly bias primordial statistical anisotropy. Indeed, upon the
inclusion of beam asymmetries into the computational scheme, the large
quadrupole statistical anisotropy vanishes. This was first shown
in Ref.~\cite{Hanson:2010gu} for the $W$ band of the WMAP7
data. To deal with this effect in the 9-year final data release,
WMAP Collaboration has produced the beam-symmetrized temperature
maps~\cite{WMAP9}.
 
This paper is organized as follows. In Sec.~2 we review vector
models of inflation predicting statistical anisotropy. We focus on the
particular class of scenarios with the Maxwellian term nonminimally
coupled to the inflaton. We review models of the (pseudo)conformal
universe in Sec.~3. In Sec.~4 we establish the estimators which
are most appropriate for our constraining purposes. In Sec.~5 we apply the
estimators to models of interest and obtain constraints on their
parameters. We compare our constraints with similar bounds obtained
from the Planck data in Sec.~6.

\section{Anisotropic inflation}

In this section we briefly discuss inflationary scenarios which 
lead to statistical anisotropy. We focus on a particular class 
of slow roll inflation that incorporates abelian gauge fields with $U(1)$
gauge symmetry. 
The healthy extension of inflation in terms of vectors 
is achieved by the following modification of the standard Maxwellian term~\cite{Watanabe:2009ct}:   
\begin{equation}
\nonumber
S_A =-\frac{1}{4}\int d^4 x \sqrt{-g} \cdot f^2 (\phi) \cdot F^{\mu \nu} F_{\mu \nu} \; .
\end{equation}
Here $f(\phi)$ is some function of the inflaton $\phi$, and $F_{\mu \nu}$ is the field strength 
of the vector field, $F_{\mu \nu}=\partial_{\mu} A_{\nu} -\partial_{\nu} A_{\mu}$. The main statement of 
Ref.~\cite{Watanabe:2009ct} is that for quite a wide range of kinetic gauge functions $f(\phi)$ 
there is a chance of obtaining a prolonged anisotropic evolution. That is, the
space-time metric tends to the attractor solution 
\begin{equation}
\label{metricanis}
ds^2 =dt^2 -e^{2Ht} \left[ e^{-4\Sigma t} dx^2 +e^{2\Sigma t} (dy^2 +dz^2)\right] \; .
\end{equation} 
Here $\Sigma$ is the parameter which measures the deviation from the rotational invariance. 
Furthermore, the evolution of inflaton fluctuations in this ``hairy'' background~\eqref{metricanis} leads 
to a directional dependence of the ACW type in the primordial power spectrum. Here we write it in the conventional form, 
\begin{equation}
\label{poweracw}
{\cal P} ({\bf k}) ={\cal P} (k) \left( 1+g_* \cos^2 \theta \right) \; , 
\end{equation}
with $\theta$ being the angle between the wave vector ${\bf k}$ and the direction $\hat{{\bf E}}_{\text{cl}}$ 
of the electric field ${\bf E}_{\text{cl}}$ generated during the inflationary stage. 
The amplitude 
$g_*$ is sourced by the electric field, i.e., $g_* \sim E^2_{\text{cl}}$, as we discuss in more detail in what follows. 
The relationship between the power spectrum~\eqref{poweracw} and the generic one~\eqref{powerangen} is given by the relations  
\begin{equation}
\label{relat}
q_{2M} = \frac{8 \pi g_*}{15} Y^{*}_{2M} (\hat{\bf E})\; , \qquad g^2_* =\frac{45}{16\pi} \sum_M |q_{2M}|^2 \; ,
\end{equation} 
which are valid in the approximation of the small amplitude $g_* \ll 1$\footnote{Upon substituting the first relation of 
Eq.~\eqref{relat} into Eq.~\eqref{powerangen}, we obtain the power spectrum ${\cal P}_{\zeta} ({\bf k})={\cal P}_{\zeta} (k) 
\left(1+g_*\cos^2 \theta -1/3 \cdot g_*\right)$, which is somewhat different from the one given by Eq.~\eqref{poweracw}. 
The difference, however, is the direction-independent piece, which can be absorbed into the redefinition of the 
spectrum ${\cal P}_{\zeta} (k)$.}. 
Note that $a(k)=1$ in Eq.~\eqref{powerangen}. 
To avoid confusion in the future, let us make one remark here. 
Though we use the terms ``electric'' and ``magnetic'', our definitions of the corresponding fields 
are different from the conventional ones. Namely~\cite{Bartolo:2012sd}, 
\begin{equation}
\nonumber
E_i =-\frac{\langle f \rangle}{a^2} A'_i, \qquad B_i =\frac{\langle f \rangle}{a^2} \epsilon_{ijk} \partial_j A_k \; ,
\end{equation}
where $\langle f \rangle$ denotes the expectation value of the
function $f(\phi)$, and a prime denotes the derivative with respect to
conformal time. 
With these definitions, the electromagnetic energy density is given by the standard expression 
$\rho_A =\frac{{\bf E}^2 +{\bf B}^2}{2}$ at all times. 

Let us discuss briefly conditions at which statistical anisotropy is generated. 
In the situation with the standard Maxwellian 
term, i.e., provided that $\langle f \rangle =1$, electric and magnetic fields 
fall down rapidly with time, as does the electromagnetic energy density $\rho_A$, which redshifts away 
as $\rho_A \propto a^{-4}$. This standard prediction can be avoided given the nontrivial structure of the 
function $f(\phi)$. A constant electric field 
is generated for the class of functions $f(\phi)$ defined by~\cite{Watanabe:2009ct} 
\begin{equation}
\nonumber
f(\phi) =e^{\frac{16c\pi}{M^2_{Pl}} \int \frac{V}{V_{\phi}} d \phi} \; , \quad c >1 \; ,
\end{equation}
where $V(\phi)$ is the inflaton potential. In Ref.~\cite{Watanabe:2009ct}, the case $c=1$ is argued to be the critical one, 
i.e., for $c<1$ the Universe 
is statistically isotropic and so is the primordial power spectrum. 
On the other hand, for $c>1$
one finds the nontrivial attractor solution 
for the electric field and the anisotropic metric of the form~\eqref{metricanis}. 
The deviation from the rotational invariance is measured by~\cite{Watanabe:2009ct, Maleknejad:2012as} 
\begin{equation}
\label{rotnoninv}
\frac{\Sigma}{H}=\frac{1}{3}\frac{c-1}{c} \epsilon \; .
\end{equation}
where $\epsilon$ is the slow-roll parameter. The amplitude of statistical anisotropy 
is related to the violation of rotational invariance, i.e., $g_* \propto 
\Sigma/H$. 

In fact, even for the case $c=1$, one finds the anisotropic 
power spectrum~\cite{Bartolo:2012sd}. This is due to the enhancement of infrared fluctuations of the gauge field by the 
standard inflationary mechanism. These quantum fluctuations 
behave like the classical field after they exit the horizon. Consequently, modes 
which exit the horizon before the time $N_{\text{CMB}} \approx 60$ in the number of $e$-folds act as an additional anisotropic background 
for inflaton fluctuations.
To account for this new effect (pointed out in Ref.~\cite{Bartolo:2012sd}), one should study the evolution 
of inflaton fluctuations in the overall classical electric field, 
\begin{equation}
\nonumber
{\bf E}_{\text{cl}} ={\bf E}_0 +{\bf E}_{\text{IR}} \; .
\end{equation}
Here ${\bf E}_0$ follows from the equation of motion (e.o.m.), while ${\bf E}_{\text{IR}}$ originates from quantum fluctuations, 
which get enhanced and classicalize 
before the time $N_{\text{CMB}}$ when inflaton perturbations leave the horizon (hence the subscript ``IR''). 
Note that ${\bf E}_{\text{IR}}$ is a random Gaussian field 
characterized by zero mean and variance~\cite{Bartolo:2012sd}, 
\begin{equation}
\label{quant}
\langle {\bf E}^2_{\text{IR}} \rangle =\frac{9H^4}{2\pi^2} N \; .
\end{equation}
Here $N$ is the number of $e$-folds from the beginning of inflation and until the time 
$N_{\text{CMB}}$, i.e., $N=N_{\text{tot}} -N_{\text{CMB}}$, with $N_{\text{tot}}$ 
being the overall number of $e$-folds during inflation. 

Taking into account both sources of statistical anisotropy, one writes the 
amplitude $g_*$ as follows~\cite{Bartolo:2012sd}: 
\begin{equation}
\label{amplitudeinfl}
g_* =-\frac{24}{\epsilon} \cdot \frac{E^2_{\text{cl}}}{V(\phi)} \cdot N^2_{\text{CMB}}\; .
\end{equation}
(See also Ref.~\cite{Lyth:2013sha}.) Quite unexpectedly, one observes a large magnitude of 
statistical anisotropy~\cite{Watanabe:2009ct, Bartolo:2012sd}. 
So, the order one amplitude $g_*={\cal O} (1)$ is obtained 
already for the ratio $\rho_A /V$ as tiny as ${\cal O} (10^{-5})$. 
The point is that one expects a much larger energy density $\rho_A$ on 
rather general grounds. Indeed, according to its quantum origin, we note that the factor $N$ in Eq.~\eqref{quant} is 
considered to be a 
large number in conventional inflationary scenarios. Hence, the naturally large amplitude $g_*$. 
To handle the situation, one assumes a tuned duration of inflation, i.e., $N_{\text{tot}} \sim N_{\text{CMB}}$. 
We reiterate this statement in the future, when imposing constraints on model parameters. 
Furthermore, the purely classical field ${\bf E}_0$ is related to the 
potential $V(\phi)$ by 
\begin{equation}
\label{electclass}
E^2_0 =\frac{c-1}{c} V(\phi) \epsilon \; .
\end{equation}
Substituting this into Eq.~\eqref{amplitudeinfl}, 
one observes that one needs extremely tuned value of the constant $c$. Namely, the order-one amplitude $g_*$ 
is obtained provided that $c-1 \sim 10^{-5}$.

To summarize, the special quadrupole predicted in the anisotropic inflation has a twofold origin 
translating into the twofold treatment of the amplitude $g_*$. If the purely classical effect 
is most relevant, 
the quantity $g_*$ is directly related to the intrinsic parameters of the model. 
There is no direct matching provided that the amplitude $g_*$ is sourced by the random 
field ${\bf E}_{\text{IR}}$. Barring fine-tuning, we focus on these two situations in what follows. 
Namely, \\ 
{\bf I} ) $E_0 \gg E_{\text{IR}}$, and, consequently, $E_{\text{cl}} \rightarrow
E_0$ is achieved in the 
formal limit $N_{\text{tot}} \rightarrow N_{\text{CMB}}$. \\ 
In this case, by substituting Eq.~\eqref{electclass} into Eq.~\eqref{amplitudeinfl}, we obtain for the amplitude  
\begin{equation}
\label{amplclass}
g_* =-24 \cdot \frac{c-1}{c} \cdot N^2_{\text{CMB}} \; .
\end{equation}
{\bf II} ) $E_0 \ll E_{\text{IR}}$, and, consequently, $E_{\text{cl}} \rightarrow E_{\text{IR}}$, 
which occurs for $c \rightarrow 1$.\\ 
We write the corresponding amplitude as follows: 
\begin{equation}
\label{quantamplinfl}
g_* =-{\bf a}^2 , \qquad {\bf a} =24 \cdot \Delta_{\zeta} \cdot N_{\text{CMB}} 
\cdot \frac{{\bf E}_{\text{IR}}}{\sqrt{2 \langle E^2_{\text{IR}}\rangle_1}} \; ,
\end{equation}
where we made use of the slow-roll relations 
\begin{equation}
\nonumber
H^2 =\frac{8\pi}{3M^2_{\text{Pl}}} V(\phi), \qquad \Delta_{\zeta} \equiv 
\sqrt{{\cal P}_{\zeta}} =\sqrt{\frac{H^4}{4\pi^2 \dot{\phi}^2}} \; , \qquad \frac{\dot{\phi}^2}{2V}=\frac{\epsilon}{3} \; .
\end{equation}
The subscript ''1'' denotes the quantity $\langle {\bf E}^2_{\text{IR}} \rangle$ formally calculated 
for the total number of $e$-folds $N_{\text{tot}}=N_{\text{CMB}}+1$. The convenience of 
the vector ${\bf a}$ introduced in Eq.~\eqref{quantamplinfl} is that its components $a_i$ obey Gaussian statistics 
(unlike the amplitude $g_*$). 
Namely, they have zero means and variances given by 
\begin{equation}
\label{vara}
\langle {\bf a}^2_i \rangle =96 \cdot {\cal P}_{\zeta} \cdot N^2_{\text{CMB}} \cdot N \; . 
\end{equation}
Considering the cases ${\bf I}$ and ${\bf II}$ separately is natural, since fields $E_0$ and $E_{\text{IR}}$ are largely unrelated 
to each other barring scenarios 
where two electric fields accidentally cancel each other with a high accuracy. With 
this qualification case ${\bf I}$ results into the constraint on the parameter $c$, 
while case ${\bf II}$ leads to an upper limit on the duration of inflation.

Equations~\eqref{relat},~\eqref{amplclass} and~\eqref{vara} will be the starting point of our discussion in Sec.~5, 
when constraining anisotropic inflation. 

\section{The (pseudo)conformal universe} 
\label{sec-1}
Generating statistical anisotropy at 
inflation requires strong assumptions 
about the inflationary stage, at least in the model we have discussed. 
On the other hand, a violation of statistical isotropy 
may arise naturally in alternative frameworks, e.g., the (pseudo)conformal universe. 
In Secs. 3.1 and 3.2, we briefly summarize some basic features inherent to this cosmological picture, 
while referring to
Refs.~\cite{Rubakov:2009np,Creminelli:2010ba,Hinterbichler:2011qk} for detailed
discussions. Predictions for statistical anisotropy are reviewed 
in Secs. 3.3 and 3.4.   

\subsection{Basic assumptions and the scale-invariant power spectrum}

In the (pseudo)conformal universe, the observed flatness 
of the primordial power spectrum is due to conformal symmetry at very 
early stages of the Universe. In more detail, there are several conditions to be satisfied~\cite{Hinterbichler:2011qk}: 
\begin{itemize} 
 \item The space-time is described by the nearly Minkowski metric at very early times.
\item The matter in the Universe is in the conformal field theory state. 
\item Among the field content of the Universe there are at least two scalars: one 
with the conformal weight $\Delta \neq 0$ and another with the weight $\Delta =0$. 
\item Classical equations of motion admit the nontrivial time-dependent solution 
of the field with $\Delta \neq 0$. 
\item The action is invariant under the shift of the weight-0 field, $\sigma \rightarrow \sigma +c$.
\end{itemize}
Given these conditions, weight-0 field perturbations evolving in the background created by the $\Delta \neq 0$ 
conformal field acquire a scale-invariant power spectrum. On the other hand, 
relaxing one or more conditions above may lead to the small scalar tilt~\cite{Hinterbichler:2011qk, Osipov:2010ee}, as required 
by the experimental data~\cite{Ade:2013zuv}.

Specifying to known realizations of the (pseudo)conformal universe, we choose the 
nonzero conformal weight to be equal to 1. Then, the generic action of the (pseudo)conformal 
universe can be written in the form   
\begin{equation}
\label{actionpseudo} 
S=S_{G+M}+S_{\rho}+\frac{1}{2} \int d^4 x \sqrt{-g} \rho^2 (\partial \sigma)^2  \; ,
\end{equation}
where $S_{G+M}$ is the action for gravity and some matter pre-existing in the early Universe. 
The third term on the 
r.h.s. describes the minimal conformal coupling of the weight-0 field $\sigma$ to the field 
$\rho$, while the second term encodes 
dynamics of the field $\rho$, which it has on its own. By assumption, the action $S_{\rho}$ allows for the 
classical solution of the field $\rho$ given by Eq.~\eqref{backg},
where the time dependence is fixed by conformal invariance. 
We recall that the parameter $h$ entering Eq.~\eqref{backg} 
is the dimensionless constant originating from the action $S_{\rho}$, while
 $t_*$ is an arbitrary constant of integration which has the meaning of the end-of-roll time.

The solution~\eqref{backg} 
spontaneously breaks conformal group $SO(4,2)$ down to the de Sitter subgroup 
$SO(4,1)$~\cite{Hinterbichler:2011qk}. 
Interestingly, this symmetry breaking pattern uniquely determines phenomenological properties of 
perturbations of the field $\sigma$. Namely, independently of the 
details of the microscopic physics, the field $\sigma$ acquires the power spectrum of the Harrison--Zel'dovich 
type~\cite{Hinterbichler:2011qk}. 
To show this explicitly, one introduces the notation $\chi =\rho_0 \delta \sigma$. From Eq.~\eqref{actionpseudo}, one derives the 
e.o.m. for the field $\chi$,
\begin{equation}
\label{eqchi}
\ddot{\chi} -\partial_i \partial_i \chi -2 h^2 \rho^2_0 \chi =0 \; .
\end{equation}
(The unit scale factor $a=1$ is assumed here). 
We observe that with the classical background $\rho_0$ as in Eq.~\eqref{backg}, the e.o.m.~\eqref{eqchi} 
coincides with that of the massless scalar field in the de Sitter background. Hence, the result 
is the same, namely, a scale-invariant power spectrum which reads in terms of 
$\sigma$ perturbations~\cite{Rubakov:2009np,Creminelli:2010ba, Hinterbichler:2011qk} 
\begin{equation}
\nonumber
{\cal P}_{\delta \sigma} =\frac{h^2}{4\pi^2} \; .
\end{equation}
We reiterate that the scale invariance as well as the results summarized
in the next subsections are largely independent of the details of the 
microscopic physics. So far, two concrete models have been
proposed. These are the conformal rolling scenario~\cite{Rubakov:2009np}
and Galilean genesis~\cite{Creminelli:2010ba}.

The former represents perhaps the simplest realization of the (pseudo)conformal universe. 
  There the field $\rho$ is a scalar 
with the standard kinetic term rolling down the negative quartic potential. 
The action for this rolling field is given by~\cite{Rubakov:2009np} 
\begin{equation}
\label{actionroll}
S_{\rho} =\int d^4 x \sqrt{-g} \left[(\partial \rho )^2+h^2 \rho^4 \right] \; .
\end{equation}
It is straightforward to show that the classical field $\rho$ has an attractor solution 
given by Eq.~\eqref{backg}. The conformal rolling scenario is natural 
from both the dynamical and spectator prospectives. In the former situation, the 
Universe driven by the field $\rho$ undergoes a slow contraction~\cite{Hinterbichler:2011qk} 
akin to ekpyrotic scenarios~\cite{Lehners:2008vx}. 
Alternatively, one treats the field $\rho$ as a spectator. In that case, the Minkowski metric 
can be imposed ``by hands''. The other possibility discussed in the 
original proposal of the conformal rolling scenario~\cite{Rubakov:2009np} is to conformally couple the spectator 
field 
$\rho$ to gravity. The background evolution of the Universe is then 
allowed to be arbitrary during the conformal phase. In what follows, we
assume the minimal coupling to gravity with minimal loss of generality.

Galilean genesis is the other example of (pseudo)conformal universe model~\cite{Creminelli:2010ba}. 
Its action is given by 
\begin{equation}
\nonumber
S_{\pi}=\int d^4 x \sqrt{-g} \left[ -f^2e^{2\pi} (\partial \pi)^2 +\frac{f^3}{\Lambda^3_G} (\partial \pi)^2 
\square \pi +\frac{f^3}{2\Lambda^3_G}(\partial \pi)^4\right] \; ,
\end{equation} 
where the field $\pi$, the Galileon, is defined by $\rho =f e^{\pi}$. 
Despite the higher derivative structure of the Galilean genesis action, the 
e.o.m. is second order in derivatives of the field $\pi$. 
Moreover, the e.o.m. admits for the solution of the form~\eqref{backg}. 
In fact, the correspondence between Galilean genesis and the conformal rolling scenario 
is much deeper: the predictions of the two models are the same modulo the replacement 
\begin{equation}
\label{duality}
h^2 \leftrightarrow \frac{2}{3}\frac{\Lambda^3_G}{f^3} \; .
\end{equation}
Note that the Galileon $\pi$ is naturally treated as a dynamical field. Driven 
by the field $\pi$, the Universe is nearly static at very early times, but slowly expands. 
Thus, the first condition outlined in the beginning of this section is satisfied 
automatically, at least at early times.

\subsection{Weight-0 perturbations: next-to-leading order}
The interaction between weight-0 and weight-1 field perturbations sources nontrivial phenomenology 
in (pseudo)conformal universe models~\cite{Libanov:2010nk, Libanov:2011hh, Creminelli:2012qr, Libanov:2011bk}. 
In particular, it gives rise to some amount of statistical anisotropy.  

Perturbations of the field $\rho$ have a red power spectrum whose form is fixed by the symmetry breaking pattern 
$SO(4,2) \rightarrow SO(4,1)$. As discussed in 
Refs.~\cite{Rubakov:2009np,Creminelli:2010ba, Hinterbichler:2011qk, Libanov:2010nk}, 
they can be absorbed into the redefinition of the end-of-roll time $t_*$, i.e.,
\begin{equation}
\label{inhomobackg}
\rho \propto \frac{1}{t_{*} ({\bf x}) -t} \; .
\end{equation}
Here
\begin{equation}
\nonumber
t_{*} ({\bf x})=t_* +\delta t_* ({\bf x}) \; ,
\end{equation}
and the shift of time $\delta t_* ({\bf x})$ is the random field with
the red power spectrum 
\begin{equation}
\nonumber
\langle \delta t^2_* ({\bf x}) \rangle \propto h^2 \int \frac{dp}{p^2} \; ,
\end{equation}
where $p=|{\bf p}|$ are wave numbers characterizing the Fourier modes of the field $\rho$. 
Clearly, the shift $\delta t_*$ as it stands is irrelevant from the physical point of view, 
since it can be absorbed into the redefinition of the end-of-roll time $t_*$.  
Interesting effects appear, once we consider the spatial variation of $t_{*} ({\bf x})$. 
It is convenient to introduce the notation 
\begin{equation}
\nonumber
v_i=-\partial_i t_{*} ({\bf x}) \; ,
\end{equation}
while keeping the standard notation for the second derivative, i.e., $\partial_i \partial_j t_*$.
The field ${\bf v}$ is a random Gaussian field with 
zero mean and variance,
\begin{equation}
\label{v2}
\langle v^2_i \rangle =\frac{3h^2}{8\pi^2} \ln \frac{H_0}{\Lambda} \; . 
\end{equation} 
Here $H_0$ is the present Hubble rate; $\Lambda$ is an infrared
cutoff. To account for the interaction with long ranged 
radial perturbations, one studies the evolution of weight-0 perturbations 
in the inhomogeneous background~\eqref{inhomobackg}. This calculation 
was done in Ref.~\cite{Libanov:2010nk} including corrections 
of the orders $\partial_i \partial_j t_{*}/k$ and $v^2$. The result for the perturbations 
$\delta \sigma$ in the late-time regime $k(t_*-t) \ll 1$ is
\begin{equation}
\label{solution}
\delta \sigma ({\bf x}, \eta)=\int \frac{d^3k}{\sqrt{k}}\frac{h}{4\pi^{3/2} \gamma (k+{\bf kv})} e^{i{\bf kx}-ikt_{*} ({\bf x})} 
\left(1-\frac{\pi}{2k} \frac{k_i k_j} {k^2} \partial_i \partial_j t_{*} +\frac{\pi}{6k} \partial_i \partial_i t_{*} \right) 
A_{\bf k} +h.c. \; .
\end{equation}
Here $\gamma =(1-v^2)^{-1/2}$, and the 
expansion to order $v^2$ is understood; $A_{\bf k}$ is the 
annihilation operator for the perturbations $\delta \sigma$. 
Note that perturbations of the weight-0 field $\sigma$ 
remain frozen out until the end of conformal rolling. 

After conformal symmetries get broken, the form of the solution~\eqref{solution} is not protected anymore. 
Still, there is the option that cosmological modes of interest are already superhorizon 
at this time, and the perturbations $\delta \sigma$ remain unchanged until the RD stage. 
In the nomenclature of Ref.~\cite{Ramazanov:2012za}, 
this is ``sub-scenario~A''. 
Note that sub-scenario~A is natural from both the dynamical and spectator prospectives. 
Alternatively, the cosmological perturbations $\delta\sigma$ may be subhorizon by the end of the roll. In that case, perturbations 
$\delta \sigma$ evolve before the beginning of the hot era. 
This option is not particularly natural in dynamical models, but can be well accomodated 
in spectator versions of the (pseudo)conformal universe. We call this
option ``sub-scenario~B''.

The two sub-scenarios lead to drastically different predictions for 
statistical anisotropy, as we discuss below.

\subsection{Sub-scenario A}
In sub-scenario A, the primordial power spectrum of scalar perturbations is derived directly from Eq.~\eqref{solution}. 
For reasons which will become clear shortly, we write it up to the quadratic order in the constant $h$, 
\begin{equation}
\label{poweranA}
{\cal P}_{\zeta} ({\bf k})={\cal P}_{\zeta} (k) \left[ 1+Q_1 ({\bf k}) + Q_2 ({\bf k})\right] \; .
\end{equation}
Here $Q_1 ({\bf k})$ is the leading order contribution, which is nonzero already in the linear 
order in $h$~\cite{Libanov:2010nk, Creminelli:2012qr}, 
\begin{equation}
\label{genquadrA}
Q_1 ({\bf k})=-\frac{\pi}{k} \hat{k}_i \hat{k}_j \partial_i \partial_j t_*  \; .
\end{equation}
This encodes a statistical anisotropy of the general quadrupole type in contrast to the 
inflationary predictions of Sec.~2. The third term in the square brackets in Eq.~\eqref{poweranA} is 
given by~\cite{Libanov:2010nk, Creminelli:2012qr}, 
\begin{equation}
\label{specquadrA}
Q_2 ({\bf k})=-\frac{3}{2} v_i v_j \hat{k}_i \hat{k}_j \; .
\end{equation}
The direction dependence present here is of the special quadrupole type akin to 
inflation with vector fields. Note its quadratic dependence on the ``velocity'', which 
implies the suppression by the additional power 
of the constant $h$ as compared to the contribution~\eqref{genquadrA}. 
This, however, does not mean that the special quadrupole can be
ignored in the 
data analysis. Moreover, if the constant $h$ is not particularly small, the term~\eqref{specquadrA} effectively 
makes 
larger imprint on the CMB sky than the general quadrupole with the amplitude decreasing 
as $k^{-1}$. At the level of cosmological measurements, the latter property
translates into the suppression at large CMB multipole number $l \propto H_0 k^{-1}$. As a result, we have
low statistics of multipoles relevant in the analysis, and hence a very weak constraint on the 
parameter $h$ from the nonobservation of the general quadrupole~\cite{Ramazanov:2012za}. 
We will return to this discussion in Sec.~5, when constraining sub-scenario A.

Let us rewrite Eq.~\eqref{genquadrA} in the conventional form, 
\begin{equation}
Q_1 ({\bf k}) =a(k) \sum_{M} q_{2M} Y_{2M} (\hat{\bf k}) \; .
\end{equation}
Here $q_{2M}$ are random Gaussian quantities with zero means and variances, 
\begin{equation}
\label{q2MA}
\langle q_{2M} q^{*}_{2M'} \rangle =\frac{\pi h^2}{25} \delta_{MM'} \; , 
\end{equation}
while the direction-independent amplitude $a(k)$ is given by 
\begin{equation}
a(k) =H_0 k^{-1} \; .
\end{equation}

On the other hand, the special quadrupole corresponding to the 
subleading-order statistical anisotropy is characterized by the scale-independent amplitude  
\begin{equation}
\label{amplitudeA}
g_* =-\frac{3}{2} v^2 \; ,
\end{equation}
and the preferred direction is associated with the unit random vector $\hat{{\bf v}} ={\bf v}/v$. 
Remarkably, this prediction is very similar
to the statistical anisotropy following from case {\bf II} of anisotropic inflation, sourced 
by the Gaussian random vector ${\bf E}_{\text{IR}}$. Formally equatting
amplitudes~\eqref{quantamplinfl} and~\eqref{amplitudeA}, and using Eqs.~\eqref{vara} and~\eqref{v2}, 
we conclude that the duality holds up to the replacement 
\begin{equation}
\label{replacement}
h^2 \ln \frac{H_0}{\Lambda} \leftrightarrow \frac{512\pi^2}{3} {\cal P}_{\zeta} N^2_{\text{CMB}} N \; .
\end{equation}
We exploit this duality when constraining sub-scenario A in Sec.~5.

\subsection{Sub-scenario B}
If cosmological modes are subhorizon by the end of the roll, they proceed to evolve at the 
so-called intermediate stage~\cite{Libanov:2011hh} which ends as modes of interest 
leave the horizon. In the conformal rolling scenario, the end of the roll is realized by 
relaxing the form of the potential in 
Eq.~\eqref{actionroll}, so that it has a minimum at some large field value $\rho=f$. 
After the field $\rho$ reaches its minimum, the 
field $\sigma$ evolves as the massless scalar minimally coupled to gravity. 
We claim that the evolution during the intermediate stage is long enough, i.e., 
$r \equiv t_1 -t_{*} \gg k^{-1}$, where $t_1$ is the time when the perturbations $\delta \sigma$ 
get frozen out (in the conventional sense). 
Second, cosmological evolution at the intermediate 
stage must be described by the nearly Minkowski metric. Otherwise, 
the flat spectrum of perturbations generated by the end of the conformal phase 
would be grossly modified. With these assumptions, the 
dynamics at the intermediate stage is fairly nontrivial, and the 
final expression for the perturbations $\delta \sigma$ 
is quite complicated. Here we simply write down the result for the power spectrum of primordial scalar 
perturbations~\cite{Libanov:2011hh},  
\begin{equation}
\nonumber
{\cal P}_{\zeta} ({\bf k})={\cal P}_{\zeta} (k) \left[ 1+{\bf n}_k ({\bf v} ({\bf n}_k r) -{\bf v} (-{\bf n}_k r) )\right] \; .
\end{equation}
Remarkably, the direction dependence present here is nonzero already at the linear order in the constant 
$h$. Moreover, it encodes the statistical anisotropy of all even multipoles starting from the quadrupole 
of the general type, i.e., all the coefficients $q_{LM}$ with even $L$ are nonzero in Eq.~\eqref{powerangen}. 
They are random Gaussian variables with zero means and variances given by~\cite{Libanov:2011hh}
\begin{equation} 
\label{qlmB}
\langle q_{LM} q^{*}_{L'M'} \rangle \equiv Q_{L} \delta_{LL'} \delta_{MM'}=\frac{3}{\pi} \frac{h^2}{(L-1) (L+2)} 
\delta_{LL'} \delta_{MM'}\; .
\end{equation}
Remarkably, the amplitude $a(k)$ does not depend on the wave number $k$, i.e., $a(k)=1$. 

The prediction of sub-scenario B for statistical anisotropy is in sharp contrast to that of inflationary scenarios.

\section{Estimators} 
As the first step, we discuss estimators for the amplitudes 
$q_{LM}$ entering the 
primordial power spectrum~\eqref{powerangen}. Following Ref.~\cite{Hanson:2009gu}, 
we make use of the QML methodology. 
This techniques was argued to be in good agreement with exact likelihood 
methods applied to the search for the statistical anisotropy with the WMAP5 data~\cite{Groeneboom:2008fz, Groeneboom:2009cb}. 
In our previous paper, we implemented QML-based estimators to the 7-year release of the WMAP and 
constrained the conformal rolling scenario from the nonobservation of statistical anisotropy. 

Let us recall the main ideas behind the QML estimator. One starts with the log-likelihood ${\cal L}$ 
of the observed sky $\hat{{\bf \Theta}}$ with respect to the 
coefficients $q_{LM}$ of statistical anisotropy. Assuming that these parameters are 
small enough, one expands the log-likelihood up to the quadratic order in $q_{LM}$'s, 
\begin{equation}
\label{logexpansion}
{\cal L} (\hat{{\bf \Theta}}| {\bf q})={\cal L}_0 +{\bf q}^{\dagger}\frac{\partial {\cal L}}{\partial {\bf q}^{\dagger}} 
\Bigl. \Bigr|_0
+\frac{1}{2} {\bf q}^{\dagger} \langle \frac{\partial^2 {\cal L}}{\partial {\bf q}^{\dagger} \partial 
{\bf q}} \rangle \Bigl. \Bigr|_0 {\bf q} \; ,
\end{equation}
where the subscript ''0'' denotes that corresponding quantities are calculated in the 
absence of statistical anisotropy. In Eq.~\eqref{logexpansion} 
we replaced the second derivative of the log-likelihood by its expectation value. 
 The QML estimator for ${\bf q}$ is obtained by setting the derivative of the quadratic 
log-likelihood to zero with respect to ${\bf q}^{\dagger}$, 
\begin{equation}
\label{qmlestim}
{\bf q} ={\bf F}^{-1}\frac{\partial {\cal L}}{\partial {\bf q}^{\dagger}}\Bigl. \Bigr|_0 \; ,
\end{equation}
Here ${\bf F}$ is the Fisher matrix defined as 
\begin{equation}
\nonumber
{\bf F} \equiv \langle \frac{\partial {\cal L}}{\partial {\bf q}} 
\frac{\partial {\cal L}}{\partial {\bf q}^{\dagger}}\rangle \Bigl. \Bigr|_0=
-\langle \frac{\partial^2 {\cal L}}{\partial {\bf q}^{\dagger} 
\partial {\bf q}}\rangle \Bigl. \Bigr|_0 \; . 
\end{equation}
The equality here follows from the normalization condition for the likelihood. 

To concretize the form of the estimator, we assume 
Gaussian temperature fluctuations. The log-likelihood then reads 
\begin{equation}
\label{Gaussianlikel}
-{\cal L} (\hat{{\bf \Theta}}|{\bf q}) \Bigl. \Bigr|_0=\frac{1}{2}\hat{{\bf \Theta}}^{\dagger} {\bf C}^{-1} \hat{{\bf \Theta}} +
\frac{1}{2} \ln \mbox{det} 
{\bf C} \; ,
\end{equation}
where ${\bf C}$ denotes the covariance matrix 
incorporating the theoretical covariance as well as the instrumental noise, ${\bf C} ={\bf S}+{\bf N}$. 
The first derivative of the log-likelihood~\eqref{Gaussianlikel} is given by 
\begin{equation}
\label{logder}
\frac{\partial {\cal L}}{\partial {\bf q}^{\dagger}} = \frac{1}{2} \bar{{\bf \Theta}}^{\dagger} 
\frac{\partial {\bf C}}{\partial {\bf q}^{\dagger}} \bar{{\bf \Theta}}-\frac{1}{2}\langle \bar{{\bf \Theta}}^{\dagger} 
\frac{\partial {\bf C}}{\partial {\bf q}^{\dagger}} \bar{{\bf \Theta}} \rangle \; .
\end{equation}
 The vector $\bar{{\bf \Theta}}$ represents the collection of 
CMB temperature coefficients filtered with the inverse isotropic covariance,  
\begin{equation}
\label{invvarfil}
\bar{{\bf \Theta}}=({\bf S}^{i}+{\bf N})^{-1} \hat{{\bf \Theta}} \; .
\end{equation}
Here ${\bf S}^{i}$ is the theoretical covariance calculated in the absence of statistical anisotropy, 
\begin{equation}
\label{diagcova}
S^{i}_{lm;l'm'}=C_l \delta_{ll'} \delta_{mm'} \; , 
\end{equation}
with $C_l$'s representing the standard angular power spectrum. 
The derivative of the covariance with respect to coefficients $q^*_{LM}$ is given by
\begin{equation}
\label{covder}
\frac{\partial C_{lm;l'm'}}{\partial q^{*}_{LM}}=i^{l'-l}C_{ll'} \int d \Omega_{{\bf k}} 
Y^{*}_{lm} (\hat{{\bf k}}) Y_{l'm'} (\hat{{\bf k}}) Y^{*}_{LM} (\hat{{\bf k}}) \; ,
\end{equation}
where
\begin{equation}
\label{cll}
C_{ll'}=4\pi  \int d \ln k \Delta_l (k) \Delta_{l'} (k) a(k) {\cal P}_{\zeta} (k) \; ,
\end{equation}
and $\Delta_l (k)$ is a transfer function. For the particular case of
$l=l'$ and 
 constant amplitude $a(k)=1$, coefficients 
$C_{ll'}$ reduce to the angular power spectrum $C_l$. Equation~\eqref{covder} 
follows from the expression for the theoretical covariance ${\bf S}$ which we write here 
for the future references
\begin{equation}
\label{theorcova}
S_{lm;l'm'}=4\pi i^{l'-l} \int \frac{d{\bf k}}{k^3} Y^{*}_{lm} (\hat{{\bf k}}) Y_{l'm'} (\hat{{\bf k}}) 
\Delta_{l} (k) \Delta_{l'} (k) {\cal P}_{\zeta} ({\bf k}) \; .
\end{equation}
It takes the diagonal form~\eqref{diagcova} 
for a statistically isotropic power spectrum ${\cal P}_{\zeta} ({\bf k})={\cal P}_{\zeta} (k)$.

The straightforward way to evaluate the Fisher matrix entering Eq.~\eqref{qmlestim} is to 
average the product of two log-likelihood derivatives over 
the large number of statistically isotropic Monte Carlo (MC) maps. A good forecast, however, is given in terms of the analytic 
Fisher matrix calculated in the homogeneous noise approximation. 
Only diagonal elements of the Fisher matrix survive in that case, i.e., 
\begin{equation} \label{fishapprox}
F_{LM;L'M'} \equiv F_L \delta_{LL'} \delta_{MM'}=\delta_{LL'} \delta_{MM'} f_{\text{sky}}\sum_{l, l'} 
\frac{(2l+1)(2l'+1)}{8\pi}\left (
\begin{array}{ccc} 
L & l & l'\\
0 & 0 & 0
\end{array} 
\right )^2 \frac{C^2_{l l'}}{C^{\text{tot}}_{l} C^{\text{tot}}_{l'}} \; ,
\end{equation}  
where $C^{\text{tot}}_{l}=C_l+N_l$; the prefactor $f_{\text{sky}}$ is an unmasked fraction of the sky. 
The formula~\eqref{fishapprox} completes the derivation of the estimators for the coefficients $q_{LM}$. 
Out of the amplitudes $q_{LM}$, one further reconstructs coefficients $C^{q}_L$ defined in the 
standard manner,
\begin{equation}
\label{cql}
C^{q}_L =\frac{1}{2L+1} \sum_{M} |q_{LM} |^2 \; . 
\end{equation}
These can be used to test the CMB statistical anisotropy in a model-independent way. 

\subsection{Statistical anisotropy of the special quadrupole type with constant amplitude}
To constrain the statistical anisotropy of the special quadrupole type, 
we slightly modify the above procedure. Our first goal is to construct 
the estimator for the amplitude $g_*$ given some fixed preferred 
direction ${\bf d}$. For this purpose, we consider the log-likelihood as the function of the unique parameter $g_*$, 
i.e., ${\cal L} (\hat{{\bf \Theta}}| g_*)$. Then, following the same steps as outlined above, we obtain  
\begin{equation}
\label{amplspecpr}
g_* =\frac{3}{2} \cdot \mbox{Re} \left( \sum_{M}  q_{2M} Y_{2M} ({\bf d}) \right)\; . 
\end{equation}
So, the estimate for the amplitude $g_*$ is reproduced immediately from estimates for the 
coefficients $q_{2M}$. Furthermore, the estimator~\eqref{amplspecpr} is unbiased and has a minimal 
variance. 

Recall that the early Universe models generically do not predict the preferred 
direction of statistical anisotropy. To estimate the amplitude 
$g_*$ in a universal way, we exploit the second relation in Eq.~\eqref{relat}, 
\begin{equation}
\label{amplnotspec}
g^2_* =\frac{45}{16\pi} \sum_M |q_{2M}|^2 \equiv \frac{225}{16 \pi} C^{q}_2 \; .
\end{equation} 
Though this estimator has an intuitively clear form, it is ``blind''
to the sign of the amplitude $g_*$. The other disadvantage of the
estimator~\eqref{amplnotspec} is that it does not discriminate
between the special and general types of quadrupoles.
Consequently, we expect somewhat weaker constraints than in the case
with the specified preferred direction. This is the price we pay for
our ignorance about the latter.

\subsection{Statistical anisotropy of the general type with Gaussian $q_{LM}$'s} 
Generically, statistical anisotropy is described by the infinite number 
of {\it random} parameters $q_{LM}$. First, we treat the case of Gaussian coefficients $q_{LM}$ as in 
sub-scenario B of the (pseudo)conformal universe. In this situation, the likelihood of the observed sky 
$\hat{{\bf \Theta}}$ is naturally considered as a function of the parameter $h^2$. The 
corresponding estimator has been derived in our previous paper~\cite{Ramazanov:2012za}. 
Let us briefly recall the main idea of the calculation. We write the likelihood of interest 
as the product of two likelihoods integrated over all possible sets of 
coefficients $\{q_{LM}\}$, i.e.,
\begin{equation} \label{totprobdens}
{\cal W} (\hat{{\bf \Theta}} | h^2)=\int {\cal W} (\hat{{\bf \Theta}}| {\bf q}){\cal W} ({\bf q}|h^2) d{\bf q} \; ,
\end{equation} 
Here ${\cal W} (\hat{{\bf \Theta}}|{\bf q})=\mbox{exp} ({\cal L})$, and ${\cal L}$ is given by Eq.~\eqref{Gaussianlikel}; 
${\cal W} ({\bf q}|h^2)$ denotes 
the likelihood of the particular realization ${\bf q}$ for a given
value of the parameter $h^2$. 
Upon using the approximation~\eqref{logexpansion}, we obtain the simple Gaussian form for the integrand 
in Eq.~\eqref{totprobdens}. Then the integral~\eqref{totprobdens} 
is evaluated in a straightforward manner. Finally, setting to zero the derivative 
of the joint likelihood with respect to the parameter $h^2$, we end up 
with the estimator~\cite{Ramazanov:2012za}
\begin{equation}
\label{estimh2}
h^2 \sum_{L} \frac{(2L+1)F^2_L \tilde{Q}^2_L}{(1+F_L \tilde{Q}_L h^2)^2}=
\sum_L \frac{(2L+1) F_L \tilde{Q}_L}{(1+F_L \tilde{Q}_L h^2)^2}
(F_L C^{q}_L -1) \; .
\end{equation}
Here the $\tilde{Q}_L$'s are defined by $Q_L =\tilde{Q}_L h^2$, and the $Q_L$'s are given by 
Eq.~\eqref{qlmB}. 
The estimator~\eqref{estimh2} is simplified considerably in the case of the quadrupole statistical anisotropy 
with the Gaussian $q_{2M}$'s,
\begin{equation}
\label{subAk}
h^2 \simeq C^{q}_2 -F^{-1}_2\; ,
\end{equation}
where we omitted the irrelevant constant prefactor. We exploit this estimator in Sec.~5 when constraining the quadrupole 
of the general type predicted in the leading order of sub-scenario A of the (pseudo)conformal universe.

Finally, let us discuss estimators for statistical anisotropy of the 
special quadrupole type governed by the {\it non-Gaussian} amplitude $g_*$. This type of 
direction dependence, we recall, arises in sub-scenario A of the (pseudo)conformal Universe 
(in the subleading order) as well as in anisotropic inflation (case {\bf II}). In that case, the discussion 
above is not applicable. Still, at the price of optimality, we choose to work with simple quadratic estimators 
built of estimators for the coefficients $q_{2M}$. Namely,
\begin{equation}
\label{constrA}
N^2,~ h^4 \ln^2 \frac{H_0}{\Lambda} \simeq C^{q}_2 \; .
\end{equation}

\section{Data analysis and results}
\begin{figure}[tb!]
\begin{center}
\includegraphics[width=0.32\columnwidth,angle=-90]{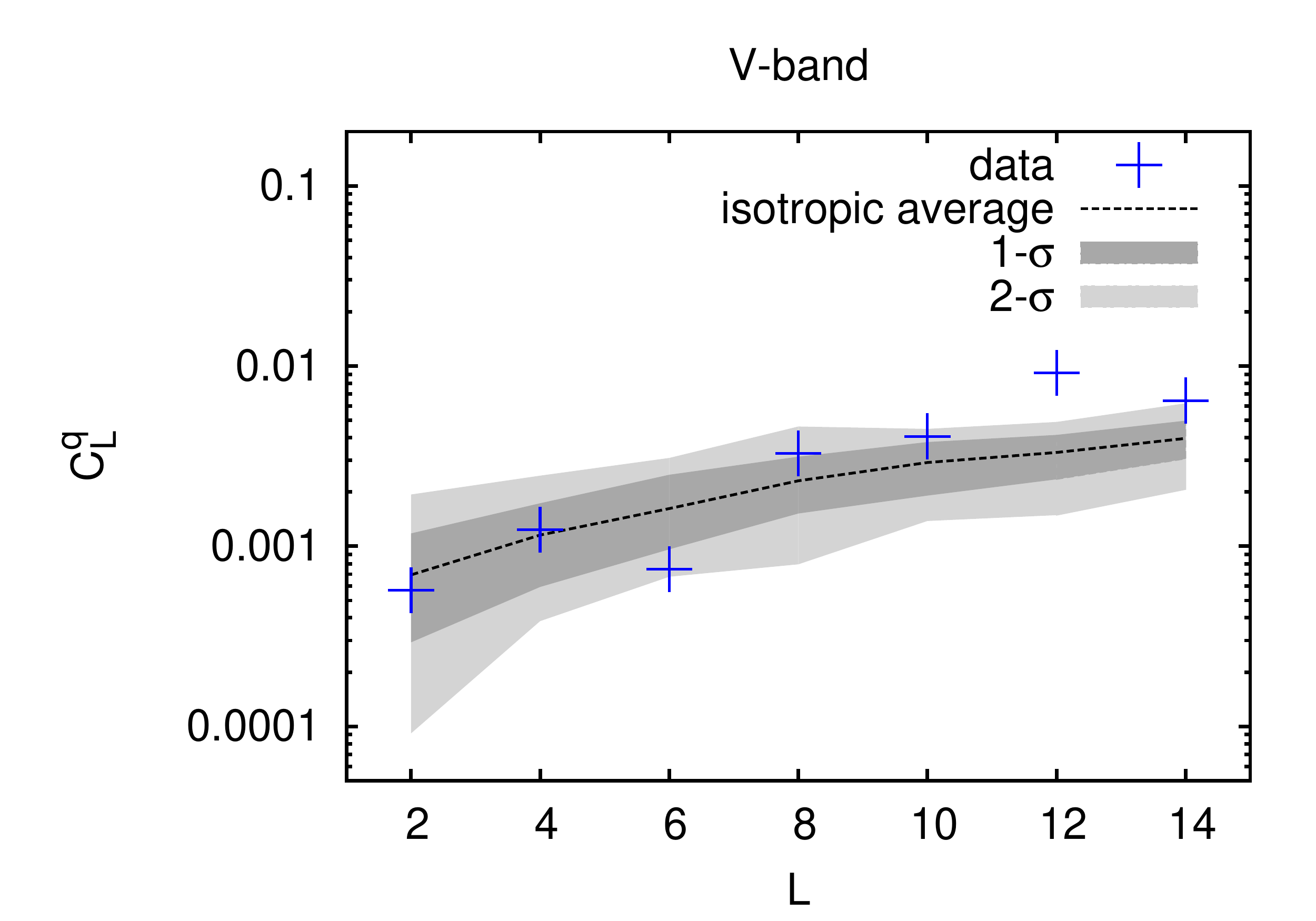}
\includegraphics[width=0.32\columnwidth,angle=-90]{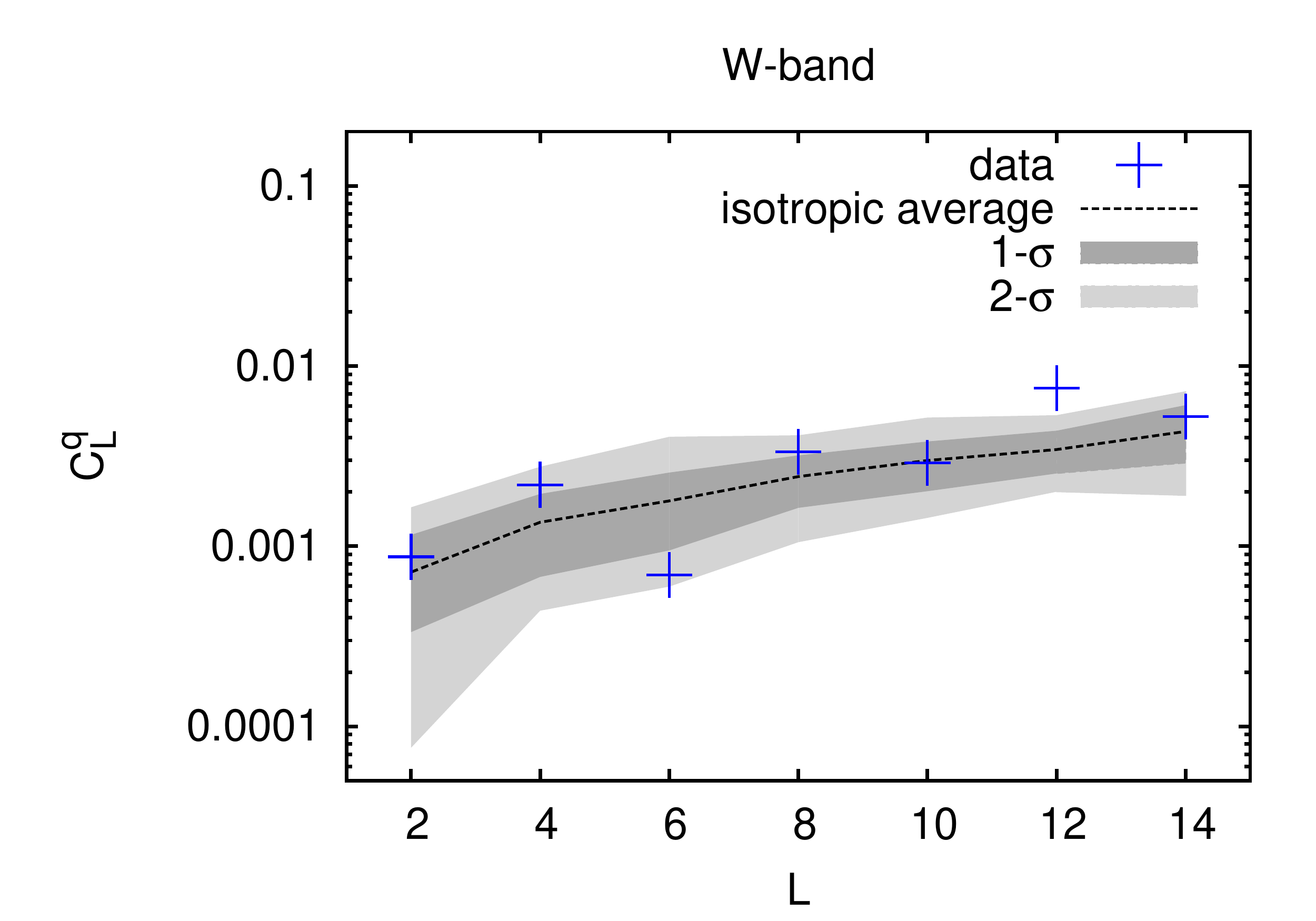}
\end{center}
\caption{The coefficients $C^{q}_L$ reconstructed from the $V$ (left) and $W$
  (right) bands of deconvolved WMAP9 maps in 
 the case of $a(k)=1$ in Eq.~\eqref{powerangen}. 
The $1\sigma$ (dark grey) and $2\sigma$ (light grey) confidence levels are overlaid 
by making use of MC-generated statistically isotropic maps. 
The analysis is done with the WMAP9 temperature analysis mask and $l_{\text{max}}=400$.}\label{modelindep}
\end{figure}

In the present section, we constrain early Universe models by
implementing estimators~\eqref{amplspecpr}, \eqref{amplnotspec},
\eqref{estimh2}, \eqref{subAk} and \eqref{constrA} to the WMAP9
data. The 9-year data set includes a new product: a set of
beam-symmetrized maps, produced by deconvolution procedure to
eliminate the effects of asymmetric beam~\cite{WMAP9}. The latter
effects were responsible for the strong bias in the measurements of
the statistical anisotropy~\cite{Hanson:2010gu}. One may treat the 
deconvolved map as a map measured by a hypothetical WMAP-like
satellite with a symmetric beam transfer function. Therefore, a simple relation between 
real and observed signals is assumed in deconvolved maps
\begin{equation}
\label{symbeams}
\hat{\Theta}_{lm} =B_l \Theta_{lm} +N_{lm} \; ,
\end{equation}
where $B_l$ is a symmetric part of the beam transfer function and $N_{lm}$
is a noise. The same relation was used in the previous
analyses~\cite{Ramazanov:2012za,Hanson:2009gu}. With this said, we can estimate coefficients $q_{LM}$ and $C^{q}_L$ literally 
following the techniques in Ref.~\cite{Hanson:2009gu}. The first and most costly step here is to
provide the inverse-variance filtering~\eqref{invvarfil}. For the
purpose of the numerical computations, we rewrite 
Eq.~\eqref{invvarfil} in the equivalent form
\begin{equation}
\label{system}
[({\bf S}^{i})^{-1}+\tilde{{\bf Y}}^{\dagger}{\bf N}^{-1}\tilde{{\bf Y}}]{\bf S}^{i} \bar{{\bf \Theta}}=
\tilde{{\bf Y}}^{\dagger} {\bf N}^{-1}  \hat{{\bf \Theta}} \; .
\end{equation}
Here $\tilde{{\bf Y}}$ is a matrix which relates harmonic space covariance and the observed map, 
\begin{equation}
\nonumber
\tilde{Y}_{ilm}=B_l Y_{lm} (i) \; .
\end{equation} 
To solve the system~\eqref{system}, we make use of the multigrid preconditioner 
proposed in Ref.~\cite{Smith:2007rg}. Having inversed-filtered temperature
anisotropies for both real data and a large number of MC maps, we evaluate estimators for the coefficients $q_{LM}$ by using 
Eq.~\eqref{qmlestim} and substituting 
Eqs.~\eqref{logder},~\eqref{covder},~\eqref{cll}, and~\eqref{fishapprox}. We compute the integral over three spherical 
harmonics in Eq.~\eqref{covder} using Slatec~\cite{slatec} and
GSL~\cite{gsl} libraries, and the coefficients $C_{ll'}$ by running
CAMB~\cite{Lewis:1999bs}. The summation over the multipole number $l$ 
in Eq.~\eqref{logder} is performed up to $l_{\text{max}}=400$. The
second term in Eq.~\eqref{logder} is calculated by avegaring over
the large number of MC maps. We evaluate the Fisher matrix using the
analytical expression~\eqref{fishapprox}. The WMAP9 \textit{kq85}
temperature analysis mask is applied to both data and MC maps leaving 
$f_{\text{sky}}=75\%$ of the sky unmasked.

In Fig.~\ref{modelindep}, we present coefficients $C^{q}_L$ reconstructed from $V$ and $W$ bands 
of the WMAP9 maps. As is clearly seen, WMAP9 data favor statistically isotropic 
primordial perturbations. 
This is to be compared with the analogous results from the 5- and 7-year releases 
revealing the anomalously large quadrupole~\cite{Hanson:2009gu, Groeneboom:2008fz, Groeneboom:2009cb}. 
Now, given the absence of the anomaly, 
we expect a substantial tightening of constraints on the early Universe models.

\begin{figure}[tb!]
\begin{center}
\includegraphics[width=0.48\columnwidth,angle=0]{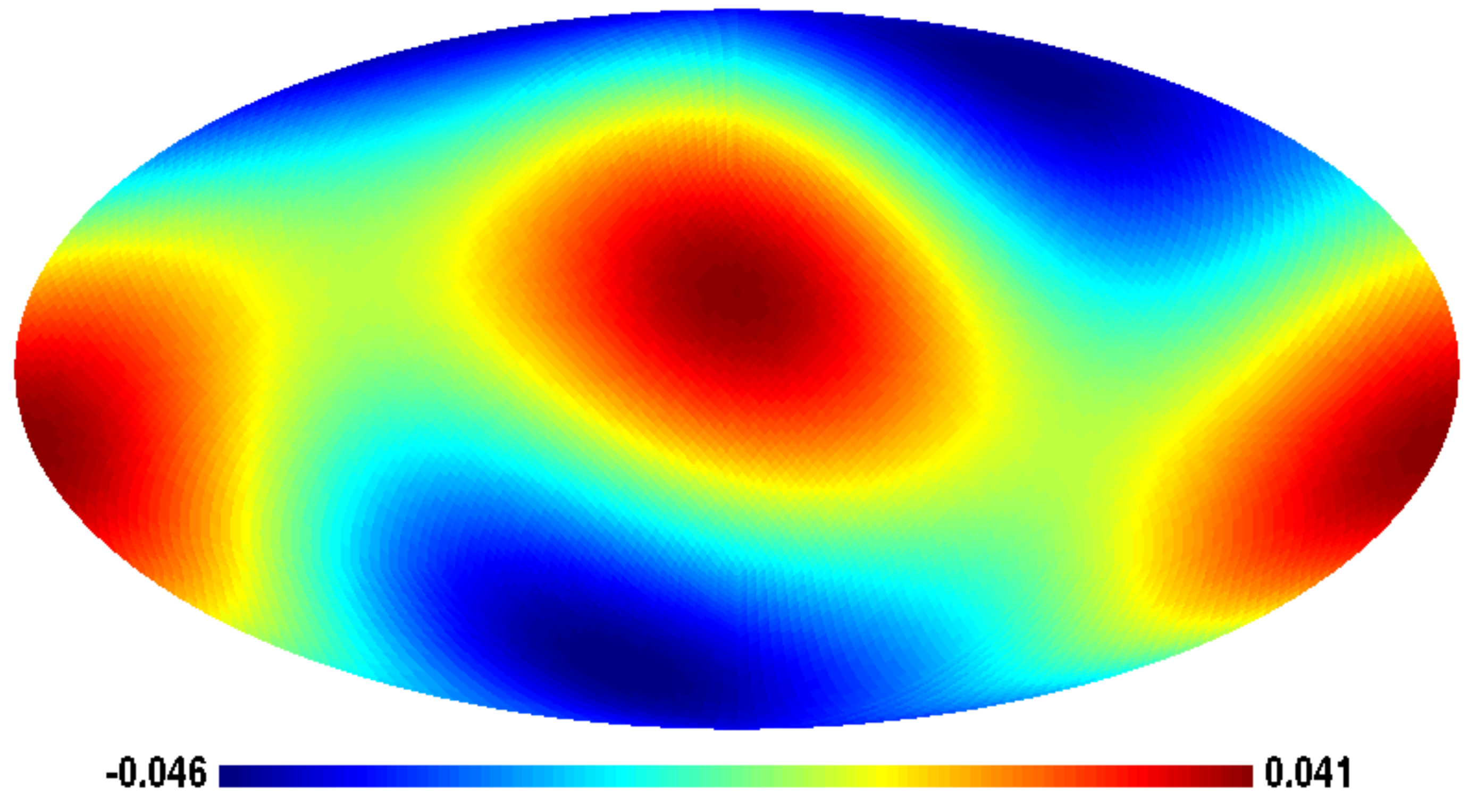}
\includegraphics[width=0.48\columnwidth,angle=0]{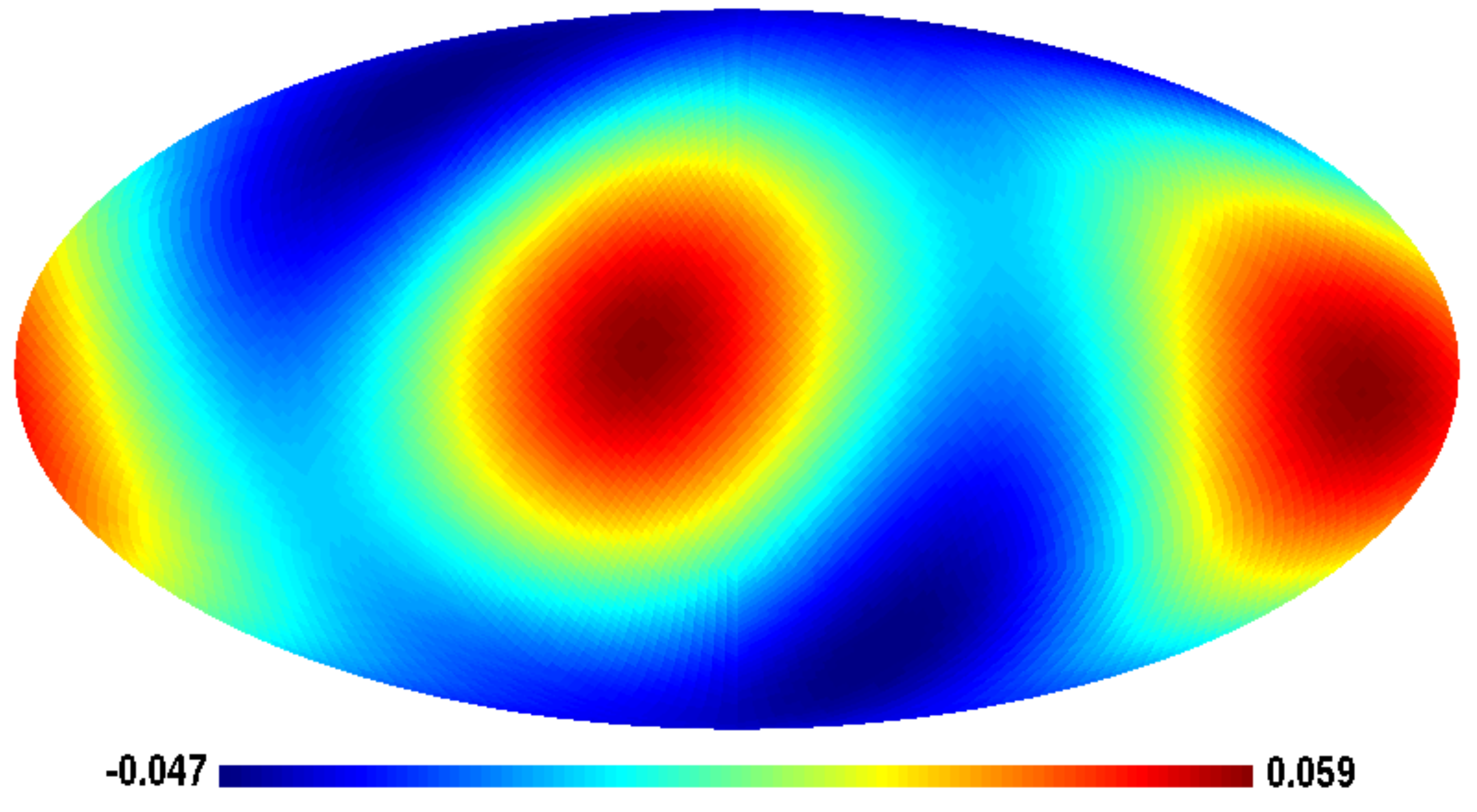}
\end{center}
\caption{The amplitude of the special quadrupole $g_*$ estimated 
from the $V$ (left) and $W$ (right) bands of deconvolved WMAP9 maps as a function of the 
direction in the sky. The plot is in the galactic coordinates with $l=180^\circ$ on the left.}\label{gstar}
\end{figure}

\subsection{Constraints on anisotropic inflation}
 
We start with constraining the amplitude $g_*$ of the quadrupole of
the special type. This is interesting from the viewpoint of scenarios
where $g_*$ is uniquely defined by the model parameters, as in the
case {\bf I} of anisotropic inflation. First, we apply the 
estimator~\eqref{amplspecpr} in order to constrain the amplitude for
some particular preferred directions. The results of the estimation
are presented in Fig.~\ref{gstar}. The constraining procedure is
as follows:
\begin{itemize}
 \item We calculate the set $\{ q_{2M}\}=q_{2,-2}, q_{2,-1},..., q_{2,2}$ 
starting from some fixed value of $g_*$ and preferred direction ${\bf d}$.  

\item For the set $\{ q_{2M}\}$, we generate a number of anisotropic maps and estimate 
the amplitude $g_*$ from the latter. We make use of Eq.~\eqref{amplspecpr}.

\item We compare the values of $g_*$ derived from anisotropic maps with the WMAP9 estimate. 
We request that not more than $95\%$ of them exceed (are smaller than) 
the real estimate in the case of positive (negative) $g_*$ fixed in the beginning. 
\end{itemize}
The generation of anisotropic maps is perhaps the most nontrivial step in this chain.
We do this in line with Appendix~A of Ref.~\cite{Hanson:2009gu} (see also 
our previous paper~\cite{Ramazanov:2012za}). The idea is to consider the temperature anisotropy of the form
 \begin{equation}
\label{anisotromaps}
{\bf \Theta}^{a}=\left({\bf I} +\delta {\bf S} [{\bf S}^{i}]^{-1}]\right)^{1/2} {\bf \Theta}^{i} \; ,
\end{equation}
where $\delta {\bf S} \equiv {\bf S} -{\bf S}^i$. Here ${\bf S}$ and ${\bf S}^i$ are the anisotropic 
and isotropic theoretical covariances given by Eqs.~\eqref{theorcova} and~\eqref{diagcova}, respectively. 
It is straightforward to check that the temperature anisotropy~\eqref{anisotromaps} corresponds to the 
anisotropic covariance ${\bf S}$. Assuming small statistical anisotropy, we expand it to the 
linear order in $\delta {\bf S}$, 
\begin{equation}
\nonumber
{\bf \Theta}^{a}={\bf \Theta}^{i} +\frac{1}{2} \delta {\bf S} [{\bf S}^i] {\bf \Theta}^i \; .
\end{equation}
We multipy the latter by the symmetric beam, convert into the pixel
space, add the noise and apply the mask. The anisotropic maps constructed 
are then analyzed in the same way as the
data maps. We followed the above procedure for the three fixed
directions in the sky. The directions form an orthogonal basis with the
first one aligned with the poles of the ecliptic plane. Constraints on the amplitude $g_*$ are presented
in Table~\ref{Constr1}. We also establish $68\%$ C.L. 
limits in the case of the direction aligned with the ecliptic poles,
\begin{equation}
\label{68c}
-0.018<g_*<0.021 \; .
\end{equation}
These limits are obtained from the $V$ band of the WMAP9 data. 
They are to be compared with the analogous constraints derived from the 
nonobservation of quadrupolar statistical anisotropy in the data by the Sloan Digital Sky Survey which read 
$g_*=0.006 \pm 0.036$ at $68\%$ C.L.~\cite{Pullen:2010zy}. 
As is clearly seen, our bounds are consistent 
with previous ones but demonstrate an improvement by a factor of 2. 

\begin{table}[htb!]
\hspace{3.2cm}
\begin{tabular}{|c|c|}
\hline 
Preferred direction & Constraint on the amplitude $g_*$\\
\hline
$(l,b)=(96.4,29.8)$ & $-0.039<g_* <0.043$ \\
\hline
$(l,b) =(96.4,60.2)$ & $-0.076<g_*<0.008$ \\
\hline
$(l,b)=(186.4,0.0)$ & $-0.022<g_*<0.078$ \\
\hline 
\end{tabular}
\caption{WMAP9 $V$ band 95\% C.L. constraints on the amplitude of the
  special quadrupole $g_*$ for particular preferred directions in the
  sky.}\label{Constr1}
\end{table}

Aiming to limit the amplitude $g_*$ without knowledge of the preferred
direction, we repeat the above procedure with minor changes. That is,
we choose about 100 random directions in the sky and calculate sets
$\{q_{2M} \}$ using Eq.~\eqref{relat}. For each set $\{ q_{2M}
\}$, we generate an anisotropic map. Now, we employ Eq.~\eqref{amplnotspec}
to estimate the strength of statistical anisotropy in real data as well as in simulated anisotropic maps. 
In fact, we construct MC maps for concrete directions, but the procedure requires that the
data and MC maps are treated on an equal footing. The results are
presented in Table~\ref{constrearly}. They demonstrate a 
substantial improvement as compared to the WMAP5 constraints, which are also presented in Table~\ref{constrearly}. As a particular application of new constraints, we set limits on
the constant $c$, a free parameter in case {\bf I} of 
anisotropic inflation. In light of the comparison
with the Planck constraints [see Eq.~\eqref{Planck}], we also establish $68\%$ C.L. 
limits, 
\begin{equation}
\label{68cl} 
-0.046 <g_* <0.048 \; ,
\end{equation}
obtained from the $V$ band of the WMAP9 data.

\begin{table}[htb!]
\hspace{-0.4cm}
\begin{tabular}{|l|c|c|c|c|}
\hline 
& 5 yr/W & 7 yr/V & 9 yr/V& 9 yr/W \\
\hline
Special quadr.
&$g_* <0.3$&- & $|g_*| <0.072$& $|g_*|<0.085$\\
\hline
Anis. infl. {\bf I} & $c-1<3.5 \cdot 10^{-6}$ & - & $c-1 <8.3\cdot 10^{-7}$ & $c-1<9.8 \cdot 10^{-7}$ \\
\hline
Anis. infl. {\bf II} &- &- & $N <82 \left(\frac{60}{N_{{\mbox CMB}}} \right)^2$& $N <128 \left(\frac{60}{N_{{\mbox CMB}}} \right)^2$\\
\hline 
Sub-sc. A (LO) &- & $h^2 < 190$ & $h^2 < 11$& $h^2<16$\\
\hline 
Gal. gen. (LO) &- &$\frac{\Lambda^3_G}{f^3} <290$ &$\frac{\Lambda^3_G}{f^3} <17$ &$\frac{\Lambda^3_G}{f^3} <24$\\
\hline 
Sub-sc. A (NLO) & -& $h^2 \ln \frac{H_0}{\Lambda} < 7$ & $h^2 \ln \frac{H_0}{\Lambda} <1.2$&$h^2 \ln \frac{H_0}{\Lambda}<2.0$\\
\hline 
Gal. gen. (NLO) &- &$\frac{\Lambda^3_G}{f^3} \mbox{ln} \frac{H_0}{\Lambda}<11$ &
 $\frac{\Lambda^3_G}{f^3} \mbox{ln} \frac{H_0}{\Lambda}<1.8$&$\frac{\Lambda^3_G}{f^3} \mbox{ln} \frac{H_0}{\Lambda}<3.0$\\
\hline 
Sub-sc. B & -& $h^2 <0.045$ & $h^2 < 0.006$& $h^2<0.013$\\
\hline
\end{tabular}
\caption{WMAP $95\%$ C.L. constraints on parameters of anisotropic models from the nonobservation 
of statistical anisotropy in the CMB sky. These include anisotropic inflation, and sub-scenarios~A and~B 
of the (pseudo)conformal universe. Constraints in the second column are nominal in a sense that they 
have been obtained by the direct comparison with the anomalous quadrupole 
as observed in Ref.~\cite{Groeneboom:2009cb}.}\label{constrearly}
\end{table}
 
The amplitude $g_*$ is random in a number of realistic models and the
constraints above are not applicable there. For example, this is the case of 
anisotropic inflation for the amplitude $g_*$ sourced by the
quantum excitations of the electric field, ${\bf E}_{\text{IR}}$. To set
limits on this class of models we slightly modify our constraining
scheme. That is, we fix some value of the $e$-fold number
$N=N_{\text{tot}}-N_{\text{CMB}}$ in the beginning. Using Eq.~\eqref{vara} with
${\cal P}_{\zeta} \approx 2.46 \cdot 10^{-9}$
we generate the collection of vectors ${\bf a}$. Each vector ${\bf
  a}$ uniquely defines the amplitude $g_*$ and the set of coefficients
$\{ q_{2M} \}$. We generate anisotropic maps for each set. Finally, we
compare the value of the $e$-fold number $N$ estimated from the
anisotropic maps with the WMAP9 estimate. We make use of the
estimator~\eqref{constrA}. Limits on the relative $e$-fold number $N
\equiv N_{\text{tot}}-N_{\text{CMB}}$ are given in Table~\ref{constrearly} at
the $95\%$ C.L. The $68\%$ C.L. limit is the strongest one for the $V$
band,
\begin{equation}
\label{68infl} 
N <14 \cdot \left( \frac{60}{N_{\text{CMB}}} \right)^2\; .
\end{equation}
A similar constraint was obtained in Ref.~\cite{Fujita:2013qxa} from the nonobservation of the trispectrum non-Gaussianity 
in the Planck data~\cite{Ade:2013ydc}; see the discussion in Sec.~6.

\begin{figure}[tb!]
\begin{center}
\includegraphics[width=0.32\columnwidth,angle=-90]{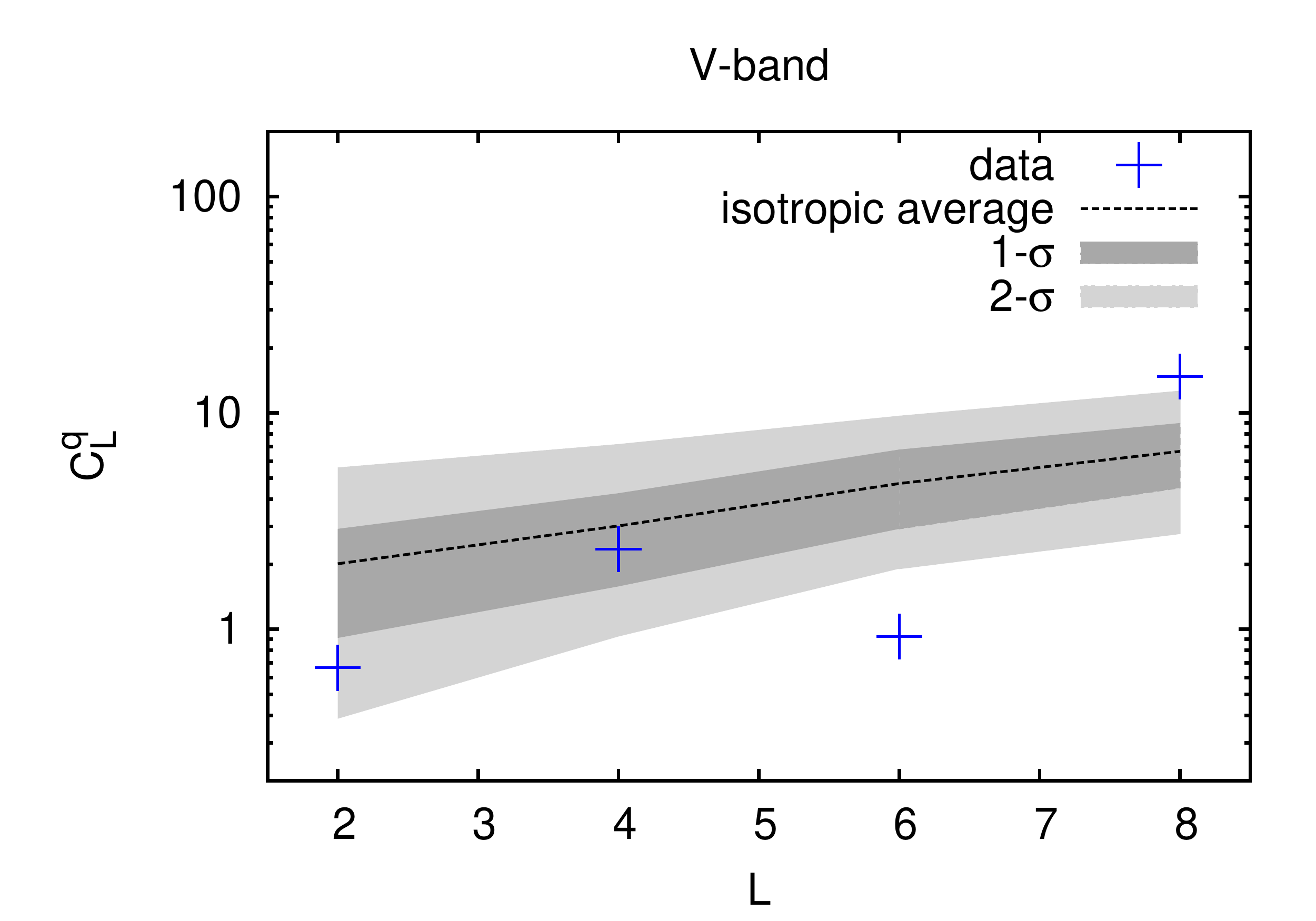}
\includegraphics[width=0.32\columnwidth,angle=-90]{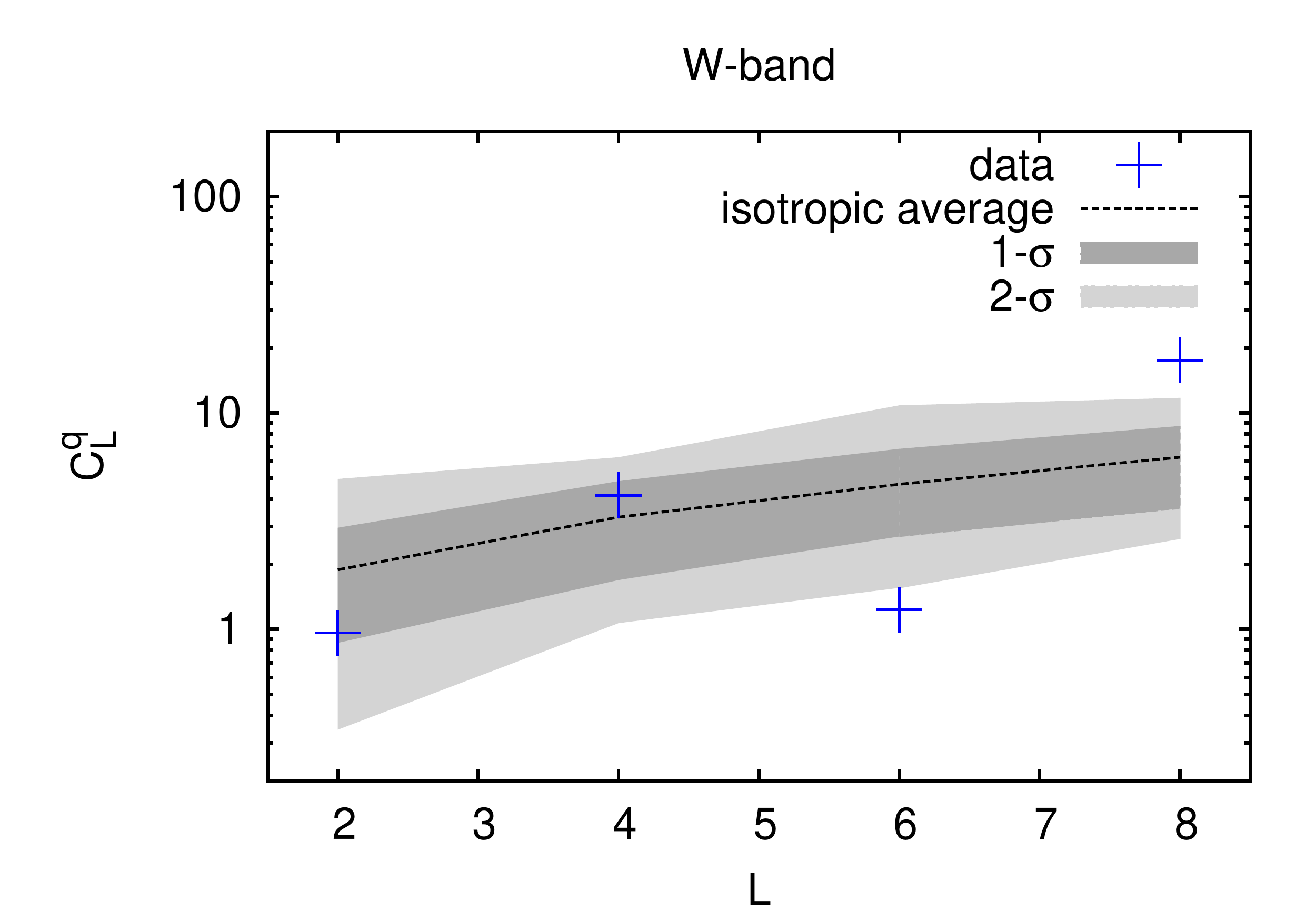}
\end{center}
\caption{The coefficients $C^{q}_L$ reconstructed from the $V$ and $W$ bands of deconvolved 
WMAP9 maps in the case of $a(k)=H_0 k^{-1}$ in Eq.~\eqref{powerangen}.
The $1\sigma$ (dark grey) and $2\sigma$ (light grey) confidence levels are overlaid 
by making use of MC generated statistically isotropic maps. The
analysis is done with the WMAP9 temperature analysis mask and $l_{\text{max}}=400$. }\label{modelk}
\end{figure}

\subsection{Constraints on the (pseudo)conformal universe}

Now, let us turn to the models of the (pseudo)conformal universe. 
We start with sub-scenario A. As discussed in Sec. 3.3, it predicts the 
quadrupolar statistical anisotropy 
of both the general and special types. The latter appears in the nonlinear order (NLO) in the 
constant $h$, while the former is nonzero already in the linear order (LO). 
Still, the general quadrupole makes a weaker imprint on the CMB sky than the special quadrupole. 
Indeed, the former is characterized by the decreasing amplitude $a(k) =H_0 k^{-1}$, 
which translates into the additional suppression by the CMB multipole 
number $l \sim k/H_0$. Consequently, we have effectively low
statistics for the multipoles useful in the analysis and, hence, a very weak
constraint on the parameter $h^2$. As a proof, the 
formal constraint $h^2< 190$ was obtained from the $V$ band of the WMAP7 data at the $95\%$ C.L. Of course, 
we cannot trust so large upper limit, since it violates the assumption of small statistical anisotropy 
made in the beginning. Rather the number ''190'' demonstrates 
the low sensitivity of the WMAP7 data towards the signal predicted. 

Qualitatively, the same story repeats at the level of WMAP9 maps. To 
show this explicitly, we estimate the coefficients $q_{LM}$ and $C^{q}_L$, but 
now with the decreasing amplitude $a(k)=H_0k^{-1}$ in
Eqs.~\eqref{powerangen} and~\eqref{cll}. Corresponding results are
shown in Fig.~\ref{modelk}. For the value of $h^2$ we use
the estimator~\eqref{subAk}. Next, we generate 100 sets of
coefficients $q_{2M}$ using Eq.~\eqref{q2MA} out of some fixed
value of the parameter $h^2$, and construct anisotropic maps for each
set. We compare values of the estimator for the parameter $h^2$
obtained from the real data and MC-simulated anisotropic maps. The final
results are presented in Table~\ref{constrearly}.

Much tighter bounds are expected from the nonobservation of
statistical anisotropy of the special type. 
Remarkably, one can derive 
them immediately from the upper limits on the relative $e$-fold number $N$ established in the 
previous subsection. This is due to the correspondence 
between predictions of two different setups, i.e., sub-scenario~A of the 
(pseudo)conformal universe and case {\bf II} of anisotropic inflation. 
The duality holds modulo the presence of the 
general quadrupole, which anyway produces the negligible effect for $h^2 \lesssim 1$. 
By exploiting Eq.~\eqref{replacement}, we obtain $95\%$ C.L. limits on the parameter of interest, 
i.e., $h^2 \ln \frac{H_0}{\Lambda}$; 
see Table~\ref{constrearly}. From Eq.~\eqref{68infl}, we also derive the $68\%$ C.L. limit,
\begin{equation}
\nonumber
h^2 \ln \frac{H_0}{\Lambda} < 0.2 \; .
\end{equation}
Finally, using the duality~\eqref{duality} between the conformal rolling scenario and Galilean genesis, 
we convert the derived constraints into the bounds on the parameter space
of the latter model; see Table~\ref{constrearly}.

\begin{figure}[tb!]
\begin{center}
\includegraphics[width=0.32\columnwidth,angle=-90]{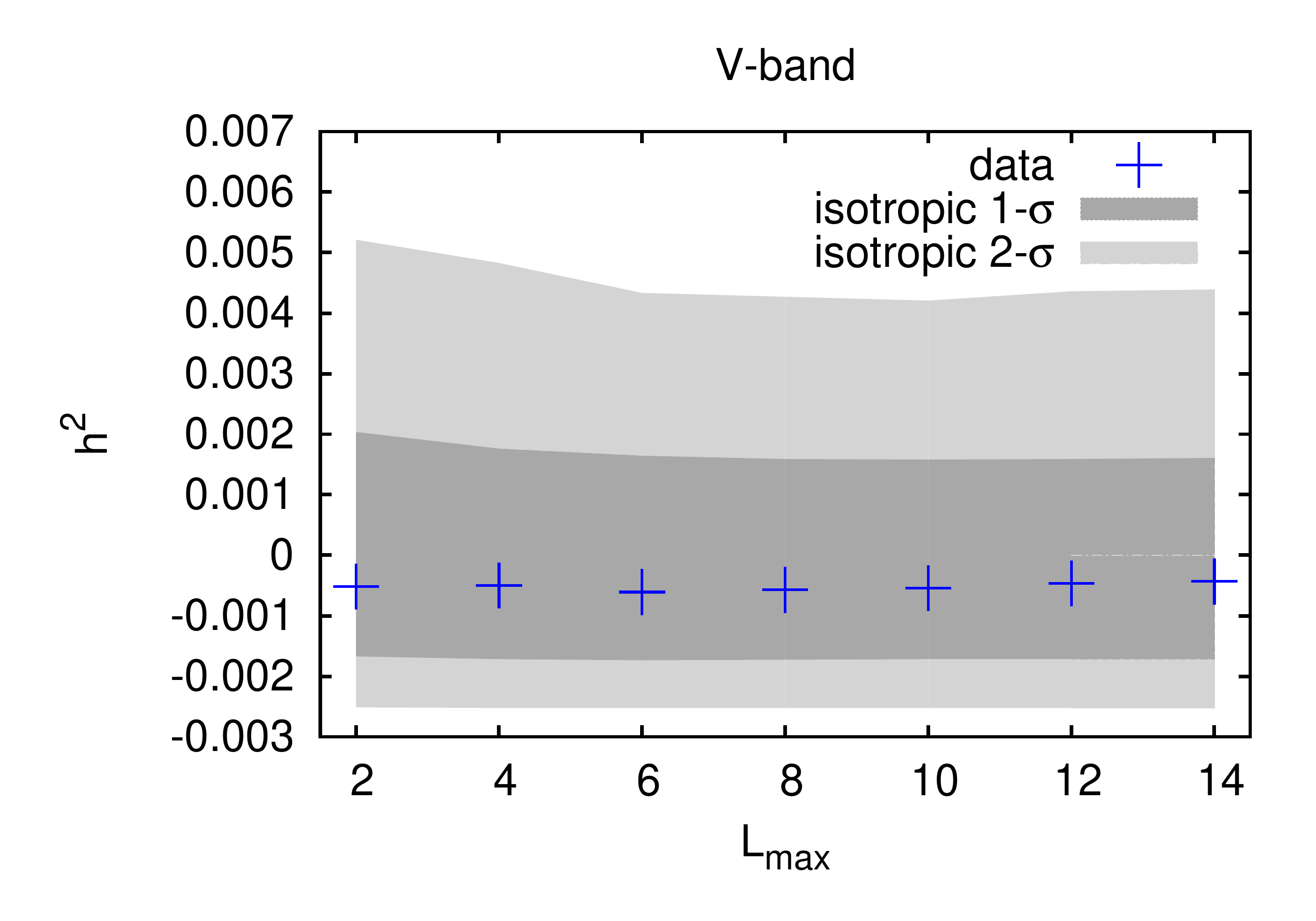}
\includegraphics[width=0.32\columnwidth,angle=-90]{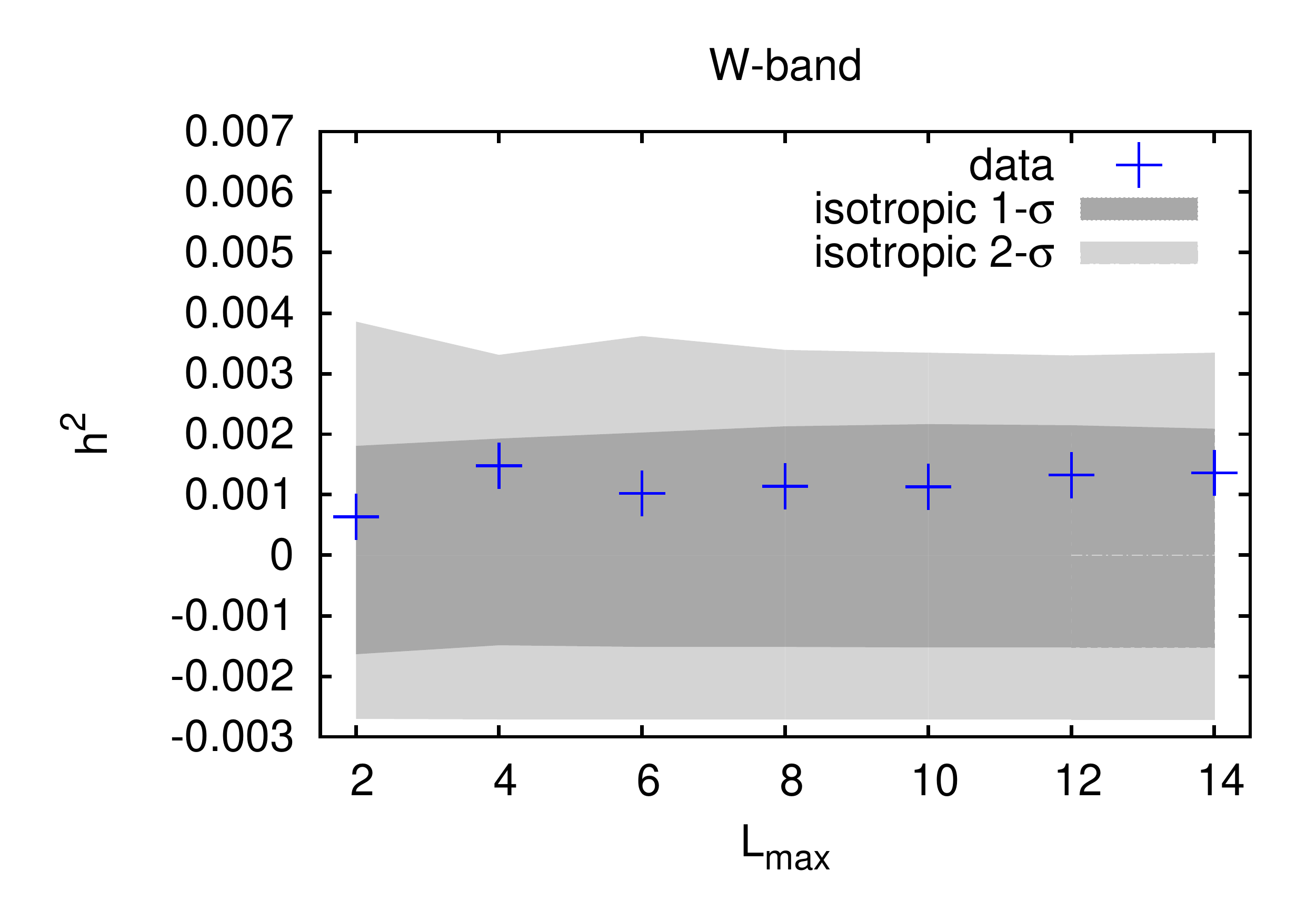}
\end{center}
\caption{Values of the estimator~\eqref{estimh2} for the 
parameter $h^2$ of the (pseudo)conformal universe, sub-scenario B, reconstructed from the $V$ (left) and $W$ (right) bands 
of deconvolved WMAP9 maps. The $1\sigma$ and $2\sigma$ confidence levels are overlaid by making use of MC-generated 
anisotropic maps.}\label{h2plot}
\end{figure}

We conclude with constraining sub-scenario~B of the (pseudo)conformal universe. 
We recall that statistically anisotropic effects in this case are nonzero already 
in the first order in the constant $h$. Given also the constant 
amplitude $a(k)=1$, we anticipate rather strong constraints from the nonobservation 
of the signal predicted. We use Eq.~\eqref{estimh2} to estimate the 
parameter $h^2$. The results plotted in Fig.~\ref{h2plot} are in excellent agreement with 
expectations from the isotropic hypothesis. Particular values of the parameter $h^2$ estimated 
at $L=14$ read $h^2=-0.0006$ ($V$ band) and $h^2=0.0007$ ($W$ band). 
To construct anisotropic maps, we generate a large number of sets of the coefficients $q_{LM}$ 
out of some fixed value of the parameter $h^2$. We make use of Eq.~\eqref{qlmB}. The constraints given in 
Table~\ref{constrearly} demonstrate a roughly one-order-of-magnitude improvement 
as compared to our WMAP7 result.

\section{Discussion}

Let us discuss our constraints in light
of the Planck data. The latter have the obvious advantage due to the 
larger number of multipoles useful in the analysis, namely 
$l^{\text{Pl}}_{\text{max}} \sim 2000$ as compared $l^{\text{W}}_{\text{max}} \sim 400$ relevant 
for the WMAP9 data. Qualitatively, this implies
\begin{equation}
\label{sensitivity}
l^{\text{Pl}}_{\text{max}}/l^{\text{W}}_{\text{max}} \sim \frac{2000}{400} =5 \; ,
\end{equation}
growth of the sensitivity to statistical 
anisotropy of the special quadrupole type. 
Indeed, the Fisher matrix scales as $l^{2}_{\text{max}}$ with the maximal 
multipole number $l_{\text{max}}$. On the other hand, the error bars for the 
amplitude $g_*$ are roughly measured by $\sqrt{F_2}$. 
This explains the estimate~\eqref{sensitivity} 
for the growth of sensitivity.
 Looking at Eq.~\eqref{68cl}, we would thus expect constraints on the
 amplitude $g_*$ at $1\%$ level in case the signal is not observed. 
Recently, $68\%$ C.L. limits appeared~\cite{kim} based on the Planck 143 GHz data at the level
\begin{equation}
\label{Planck}
-0.014<g_*< 0.018 \; ,
\end{equation}
independently of the preferred direction in the sky. These Planck constraints are stronger than the 
limits~\eqref{68cl} derived 
from the WMAP9 data by a factor 2-3. 
The first Planck limits were obtained, however, using only one
frequency band and a suboptimal estimator. 

Based on the above, we anticipate somewhat stronger
constraints with the future analysis of Planck data. Interesting
consequences are expected for the duration of inflation in models
with the nonminimal Maxwellian term: in particular, we expect 
extremely tuned number of $e$-folds in those models, i.e., $N_{\text{tot}}-N_{\text{CMB}}
\simeq {\cal O} (1)$. Planck data are also promising for constraining 
sub-scenario~B of the (pseudo)conformal universe.  Indeed, the
sensitivity of the data to the parameter $h^2$ grows as
$l^2_{\text{max}}$. Qualitatively, this implies a bound on
$h^2$ that is stronger by a factor of 
$\left(l^{\text{Pl}}_{\text{max}}/l^{\text{W}}_{\text{max}} \right)^2 \sim 25$ in the case where the signal of interest is not
observed. A much weaker improvement, i.e., by a factor
$l^{\text{Pl}}_{\text{max}}/l^{\text{W}}_{\text{max}} \sim 5$, is expected for the parameter $h^2
\ln \frac{H_0}{\Lambda}$ of sub-scenario~A.

We end up with a few remarks. The types of statistical
anisotropy studied in this paper cover most predictions that 
exist in the literature, but not all. For example,
the model of Ref.~\cite{Urban:2013spa} predicts a special
quadrupole characterized by an increasing amplitude, i.e., $g_* \sim
k^4$. This originates from the axial coupling between the inflaton and
vector fields. The peculiar form of statistical anisotropy with a vanishing
quadrupole term follows from inflation involving scalars with
nonminimal kinetic terms~\cite{ArmendarizPicon:2007nr}. In both
cases, a separate data analysis is required. On the other hand,
constraints presented in Table~\ref{constrearly} are easily
converted into limits on the parameters of scenarios with more
conventional predictions, e.g., statistical anisotropy of the ACW
type. These include scenarios based on the noncommutative geometry~\cite{DiGrezia:2003ug}, $p$-forms~\cite{Germani:2009iq}, etc.

Although we focused on the particular prediction of statistical
anisotropy, the scenarios discussed in this paper may have other
interesting signatures. Indeed, anisotropic inflation gives rise to
some amount of non-Gaussianities at both the bispectrum~\cite{Bartolo:2012sd, Abolhasani:2013zya} and trispectrum~\cite{Fujita:2013qxa} 
levels. These are sourced by infrared fluctuations of the electric
field, and thus are highly sensitive to the duration of the inflationary
phase. In Ref.~\cite{Fujita:2013qxa}, this simple observation has
been used to establish the upper limit
\begin{equation}
\label{ngconstr}
N \lesssim 17 \times \left(\frac{N_{\text{CMB}}}{50} \right)^4 \cdot \left(\frac{\tau_{\text{NL}}}{2800} \right)\; .
\end{equation}
Also given that the Planck constraint for the trispectrum parameter
 $\tau_{\text{NL}} <2800$ at $95\%$ C.L., one concludes that
the constraint~\eqref{ngconstr} is fairly similar to our $68\%$
C.L. limit~\eqref{68infl}. Note, however, that the random nature of the
field ${\bf E}_{\text{IR}}$ was not accounted for in the derivation of
Eq.~\eqref{ngconstr}. Thus, the latter is less conservative by its 
definition and may become weaker if the
randomness is included.
 Other predictions of inflation with the nonminimal
Maxwellian term include the anisotropy in the tensor power spectrum
and the cross correlation between curvature and tensor perturbations~\cite{Gumrukcuoglu:2010yc, Soda:2012zm}.

Models of the (pseudo)conformal universe may also lead to potentially large non-Gaussianities
~\cite{Creminelli:2012qr, Libanov:2011bk, Hinterbichler:2012mv}. In particular, the 
trispectrum~\cite{Creminelli:2012qr, Libanov:2011bk} governed by the parameter $h^2$, can be used to 
constrain the latter from nonobservation in the Planck data. 
This approach appears to be especially promising in a view of the fact
that sub-scenario~A predicts rather weak signal of statistical anisotropy.

\section*{Acknowledgments} 

It is a pleasure to thank Valery Rubakov, Diego Chialva, Alessandro Gruppuso, Maxim Libanov, and Federico Urban 
for many useful comments and stimulating discussions. 
We are grateful to D.~Hanson for kindly providing the code for inverse variance filtering. 
S.~R. is supported by the Belgian Science Policy (IAP VII/37). S.~R. is indebted to Lund University 
for warm hospitality. The work of G.~R. is supported in part by the RFBR grants 
12-02-00653, 12-02-31776, 13-02-01293, by the Dynasty Foundation, by the grants of the President 
of the Russian Federation NS-5590.2012.2, MK-1170.2013.2, by the Russian Federation Governement 
Grant No.11.G34.31.0047.  We acknowledge 
the use of the Legacy Archive for Microwave Background Data Analysis (LAMBDA), part of the High Energy Astrophysics 
Science Archive Center (HEASARC). HEASARC is a service of the Astrophysics Science Division at the NASA Goddard 
Space Flight Center. The numerical part of the work was done at the cluster of the Theoretical Division of INR RAS.

\end{document}